\documentclass{ieeeaccess}
\usepackage{cite}
\usepackage{amsmath,amssymb,amsfonts}
\usepackage{algorithmic}
\usepackage{graphicx}
\usepackage{multicol}
\usepackage[nolist]{acronym}
\usepackage{tablefootnote}
\usepackage{booktabs}
\usepackage{amssymb}%
\usepackage[utf8]{inputenc}
\usepackage{array}
\usepackage{url}
\usepackage{balance}
\usepackage{academicons}

\usepackage{subcaption}
\usepackage{pifont}% http://ctan.org/pkg/pifont
\newcommand{\cmark}{\ding{51}}%
\newcommand{\xmark}{\text{\ding{55}}}
\usepackage{textcomp}
\def\BibTeX{{\rm B\kern-.05em{\sc i\kern-.025em b}\kern-.08em
    T\kern-.1667em\lower.7ex\hbox{E}\kern-.125emX}}

\usepackage{hyperref} %<--- Load after everything else

\begin{document}
\history{Date of publication xxxx 00, 0000, date of current version xxxx 00, 0000.}
\doi{10.1109/ACCESS.2017.DOI}

\title{Artificial Intelligence for Cochlear Implants: Review of Strategies, Challenges, and Perspectives}
\author{Billel Essaid\authorrefmark{1}, Hamza Kheddar\authorrefmark{1},\IEEEmembership{Senior Member, IEEE}, Noureddine Batel, \authorrefmark{1}, Muhammad E. H. Chowdhury \authorrefmark{2},  \IEEEmembership{Senior Member, IEEE}, and  
Abderrahmane Lakas, \IEEEmembership{Senior Member, IEEE}\authorrefmark{3}}

\address[1]{LSEA Laboratory, Electrical Engineering Department, University of Medea, 26000, Algeria.}

\address[2]{Department of Electrical Engineering, Qatar University, Doha 2713, Qatar.}

\address[3]{College of Information Technology, United Arab Emirates University, Al Ain P.O. Box 17551, United Arab Emirates.}

\markboth
{Author \headeretal: Preparation of Papers for IEEE TRANSACTIONS and JOURNALS}
{Author \headeretal: Preparation of Papers for IEEE TRANSACTIONS and JOURNALS}

\corresp{Corresponding author: Muhammad E. H. Chowdhury (e-mail: mchowdhury@qu.edu.qa)}

\begin{abstract}
Automatic speech recognition (ASR) plays a pivotal role in our daily lives, offering utility not only for interacting with machines but also for facilitating communication for individuals with partial or profound hearing impairments. The process involves receiving the speech signal in analog form, followed by various signal processing algorithms to make it compatible with devices of limited capacities, such as cochlear implants (CIs). Unfortunately, these implants, equipped with a finite number of electrodes, often result in speech distortion during synthesis. Despite efforts by researchers to enhance received speech quality using various \ac{SOTA} signal processing techniques, challenges persist, especially in scenarios involving multiple sources of speech, environmental noise, and other adverse conditions. The advent of new artificial intelligence (AI) methods has ushered in cutting-edge strategies to address the limitations and difficulties associated with traditional signal processing techniques dedicated to CIs. This review aims to comprehensively cover advancements in CI-based ASR and speech enhancement, among other related aspects. The primary objective is to provide a thorough overview of metrics and datasets, exploring the capabilities of AI algorithms in this biomedical field, and summarizing and commenting on the best results obtained. Additionally, the review will delve into potential applications and suggest future directions to bridge existing research gaps in this domain.
\end{abstract}


\titlepgskip=-15pt

\maketitle

\begin{table*}[]
    \centering
%{\small \section*{Acronyms and Abbreviations}}
\begin{multicols}{3}
\footnotesize
\begin{acronym}[GCPFE]
\acro{CNN}{convolutional neural network}
\acro{AI}{artificial intelligent}
\acro{DL}{deep learning}
\acro{CI}{cochlear implant}
\acro{ML}{machine learning}
\acro{FCN}{fully convolutional neural networks}
\acro{STOI}{short time  objective intelligibility}
\acro{MMSE}{minimum mean square error}
\acro{DDAE}{deep denoising auto-encoder}
\acro{SNR}{signal to noise ratio}
\acro{EAS}{electric and acoustic stimulation}
\acro{CGRU}{ convolutional recurrent neural network with gated recurrent units}
\acro{CWT}{ continuous wavelet transform}
\acro{GRU}{gated recurrent units}
\acro{IGCIP}{image guided cochlear implant programming }
\acro{ESTOI}{extended short-time
objective intelligibility}
\acro{Bi-LSTM}{bidirectional long short term memory }
\acro{AAD}{auditory attention decoding}
\acro{SVM}{support vector machine}
\acro{CT}{computed tomography}
\acro{OCT}{optical coherence tomography}
\acro{MEE}{mean endpoint error}
\acro{TMHINT}{taiwan mandarin hearing in noise test}
\acro{Acc}{Accuracy}
\acro{Sen}{Sensitivity}
\acro{Spe}{Specificity}
\acro{Pre}{Precision}
\acro{Rec}{Recall}
\acro{F1}{F1 score}
\acro{NCM}{normalized covariance measure}
\acro{FFT}{fast Fourier transform}
\acro{ACE}{advanced combination encoder}
\acro{LGF}{loudness growth function}
\acro{MMS}{min-max similarity}
\acro{BF}{boundary F1}
\acro{ROC}{ receiver operating characteristic curve}
\acro{NCM}{normalized covariance measure}
\acro{DCS}{dice coefficient similarity}
\acro{ASD}{average surface distance}
\acro{AVD}{average volume difference}
\acro{JI}{jaccard index}
\acro{IoU}{intersection over union}
\acro{MAE}{mean absolute error}
\acro{HD}{hausdorff distance }
\acro{VAE}{variational autoencoders}
\acro{CAE}{convolutional autoencoder}
\acro{AE}{autoencoder}
\acro{SOTA}{state-of-the-art}
\acro{MFCC}{mel-frequency cepstral coefficient}
\acro{GFCC}{gammatone frequency cepstral coefficient }
\acro{AMR}{adaptive multi-rate}
\acro{LSTM}{long short term memory}
\acro{ASR}{automatic speech recognition}
\acro{DNN}{deep neural networks}
\acro{CS}{channel selection}
\acro{RNN}{recurrent neural network}
\acro{GAN}{generative adversarial network}
\acro{PESQ}{ perceptual estimation of speech quality}
\acro{KLT}{karhunen-loéve transform}
\acro{logMMSE}{log minimum mean squared error}
\acro{DAE}{deep autoencoder}
\acro{ICA}{intra cochlear anatomy}
\acro{ST}{scala tympani}
\acro{SV}{scala vestibul}
\acro{AR}{active region}
\acro{ASM}{active shape model}
\acro{NR}{noise reduction}
\acro{Res-CA}{residual channel attention}
\acro{GCPFE}{global context-aware pyramid feature extraction}
\acro{ACE-Loss}{active contour with elastic loss}
\acro{DS}{deep supervision}
\acro{UHR}{ultra-high-resolution}
\acro{ESCSO}{enhanced swarm-based crow search optimization}
\acro{cGAN}{conditional generative adversarial networks}
\acro{MAR}{metal artifacts reduction}
\acro{P2PE}{point to point error}
\acro{ASE}{average surface error}
\acro{MARGAN}{metal artifact reduction based generative adversarial networks}
\acro{ESCSO}{ enhanced swarm based crow search optimization}
\acro{EEG}{electroencephalography}
\acro{ERP}{event-related potential}
\acro{ANN}{artificial neural network}
\acro{RBNN}{radial basis functions neural networks}
\acro{KNN}{K-nearest neighbours}
\acro{RF}{random forests}
\acro{DRNN}{deep recurrent neural networks}
\acro{MLP}{multilayer perceptrons}
\acro{NMF}{non-negative matrix factorization}
\acro{DCAE}{deep convolutional auto-encoders}
\acro{ILD}{interaural level difference}
\acro{RTF}{relative transfer function}
\acro{SDR}{source-to-distortion ratio}
\acro{SAR}{source-to-artifact ratio}
\acro{SIR}{source-to-interference ratio}
\acro{dB}{decibel}
\acro{MIMO}{multiple-input multiple-output}
\acro{EVD}{eigenvector decomposition}
\acro{ReLU}{rectified linear unit}
\acro{RNN}{recurrent neural network}
\acro{MICE}{multiple imputation by chained equations}
\acro{ECochG}{intracochlear electrocochleography}
\acro{CM}{cochlear microphonic}
\acro{FFR}{frequency following responses}
\acro{$f_0$}{fundamental frequency}
\acro{TFSC}{temporal fine structure cues}
\acro{RMSE}{root mean square error}
\acro{BM}{basilar-membrane}
\acro{ROI}{regions of interest}
\acro{DRL}{deep reinforcement learning}
\acro{RL}{reinforcement learning}
\acro{ABR}{auditory brainstem responses}
\acro{DBS}{deep brain stimulation}
\acro{FOX}{fitting to outcomes expert}
\acro{DSP}{digital signal processor}
\acro{MHINT}{mandarin of hearing in noise 
test}
\acro{BCP}{bern cocktail party}
\acro{EST}{effective stimulation threshold}
\acro{ECAP}{electrically evoked compound action potential}
\acro{MRI}{magnetic resonance 
imaging}
\acro{BSS}{blind source separation}
\acro{AEP}{auditory evoked potential}
\acro{CCA}{canonical correlation analysis}
\acro{ASE}{average surface error}
\acro{STOI}{short-time objective intelligibility}
\acro{TL}{transfer learning}
\acro{CIS}{continuous interleaved sampling}
\acro{DTL}{deep transfer learning}
\acro{DT}{decision tree}
\acro{FL}{federated learning}
\acro{CTC}{connectionist temporal classification}
\acro{BERT}{bidirectional encoder representations from transformer}
\end{acronym}

\end{multicols}
\end{table*}

\section{Introduction}
\label{sec1}

In the symphony of modern technology, \ac{ASR} emerges as a master, orchestrating a seamless interaction between humans and machines. This transformative technology has quietly become an integral part of our daily lives, influencing how we communicate, access information, and even navigate the intricacies of healthcare. The significance of \ac{ASR} extends beyond its role in facilitating human-computer interaction; it permeates diverse applications such as voice assistants and virtual agents, speech-to-text conversion, identity verification, and holds particular promise in the realm of biomedical research \cite{er2021parkinson,KheddarASR2023}. \ac{ASR} bridges the gap between spoken language and digital communication, enabling the conversion of spoken words into written text with remarkable accuracy. The pervasiveness of \ac{ASR} technology is evident in the devices we use daily—smartphones, smart speakers, and voice-activated virtual assistants—all seamlessly responding to our spoken commands and queries. The convenience it brings to our lives is undeniable, offering a hands-free and efficient mode of interaction that has become second nature.

\ac{ASR} is pivotal in authentication systems, safeguarding the security and privacy of sensitive information. The integrity of audio speech can be verified through techniques-based \ac{ASR} such as adversarial attack detection \cite{hu2019adversarial}, steganalysis \cite{kheddar2022high,kheddar2023deep,kheddar2018fourier}, speech biometrics \cite{singh2018voice}, and more. Beyond the realm of communication, \ac{ASR} finds itself at the heart of various applications, each playing a unique role in different domains. \textit{Speaker recognition}, a facet of \ac{ASR}, is not merely confined to enhancing security measures. It has evolved into a versatile tool employed in healthcare, where the identification of individuals through their unique vocal signatures holds promise for personalized patient care. This is particularly relevant in scenarios where quick and secure authentication is crucial, such as accessing medical records or authorizing medical procedures. \textit{Event recognition}, another dimension of \ac{ASR}, is a game-changer in sectors ranging from security to healthcare. In the former, \ac{ASR} algorithms analyze audio data to automatically detect and categorize specific events, reinforcing surveillance capabilities. In healthcare, event recognition becomes a powerful tool for monitoring and early detection of health-related events, recognizing speech from noisy environment \cite{djeffal2023noise}, understanding person with dysarthria severity \cite{hamza2023machine}, and more. In the context of cardiac health, \ac{ASR} can aid in identifying anomalies in heart sounds, potentially enabling early intervention and preventive measures \cite{chen2016s1}. \textit{Source separation}, the ability to discern and isolate individual sound sources from complex audio signals, is a boon in fields like entertainment and music production. However, its significance extends into the realm of biomedical research, where \ac{ASR} plays a pivotal role in decoding the intricate language of physiological signals. In the context of \acp{CI}, source separation becomes a critical component in enhancing the auditory experience for individuals with hearing impairments.

The \acp{CI}, designed to restore hearing in individuals with severe hearing loss or deafness, rely on \ac{ASR} for optimizing their functionality. \ac{ASR} contributes significantly to the improvement of speech perception in \ac{CI} users by enhancing the processing and interpretation of auditory signals. \acp{CI} work by converting sound waves into electrical signals that stimulate the auditory nerve, bypassing damaged parts of the inner ear. \ac{ASR} complements this process by aiding in the recognition and translation of spoken language. The technology plays a crucial role in optimizing speech understanding for \ac{CI} users by refining the interpretation of varied speech patterns, tones, and nuances. Moreover, \ac{ASR} in the context of \acp{CI} extends beyond basic speech recognition. It contributes to the recognition of environmental sounds, facilitating a more immersive auditory experience for individuals with hearing impairments. This is particularly significant in enhancing the quality of life for \ac{CI} recipients, allowing them to navigate and engage with their surroundings more effectively.

\subsection{Related work}

Many reviews have been written in the context of \ac{CI}. For example, \cite{crowson2020machine} discussed the advantages offered by \ac{ML} to cochlear implantation, such as analyzing data to personalize treatment strategies. It enhances accuracy in speech processing optimization, surgical anatomy location prediction, and electrode placement discrimination. Besides, it delves into its applications, including optimizing cochlear implant fitting, predicting patient threshold levels, and automating image-guided CI surgery. The review discusses some novel opportunities for research, emphasizing the need for high-quality data inputs and addressing concerns about algorithm transparency in clinical decision-making for improved patient care. Similarly, the review by Manero et al. \cite{manero2023improving} details some benefits of employing \ac{AI} in enhancing \ac{CI} technology, involving adaptive sound processing, acoustic scene classification, and auditory scene analysis. The authors discuss AI-driven advancements aiming to optimize sound signals, adapt to diverse environments, and improve speech perception for individuals with hearing loss, ultimately enhancing their overall quality of life. 

Additionally, the review \cite{wilson2022harnessing} explores three main topics: direct-speech neuroprosthesis, which involves decoding speech from the sensorimotor cortex using \ac{AI}, including the synthesis of produced speech from brain activity; a top-down exploration of pediatric cochlear implantation using \ac{ML}, delving into its applications in pediatric cochlear implantation; and the potential of \ac{AI} to solve the hearing-in-noise problem, examining its capabilities in addressing challenges related to hearing in noisy environments. Moreover, the review \cite{d2022tele} critically examines the current landscape of tele-audiology practices, highlighting both their constraints and potential opportunities. Specifically, it explores intervention and rehabilitation efforts for \acp{CI}, focusing on remote programming and the concept of self-fitting \acp{CI}. Recently, a review by Henry et al. in 2023 \cite{henry2021noise} conducts a comprehensive review of noise reduction algorithms employed in \acp{CI}. Maintaining a general classification based on the number of microphones used—single or multiple channels—the analysis extends to incorporate recent studies showcasing a growing interest in \ac{ML} techniques. The review culminates with an exploration of potential research avenues that hold promise for future advancements in the field. Table \ref{tab:cmp} offers a comparative analysis of the proposed review in contrast to other discussed AI-based \ac{CI} reviews and surveys.

\begin{table*}[t!]
\centering
\caption{Comparison of this review against other existing AI-based \ac{CI} reviews and surveys. Tick marks (\cmark) indicate that a particular field has been considered, while cross marks (\xmark) indicate that a field has been left unaddressed. The symbol (\ding{72}) indicates that the most critical concerns of a field have not been addressed.}
\label{tab:cmp}

\begin{tabular}{m{0.5cm}m{0.5cm}m{4.5cm}ccccccccc}
\hline
Ref.  & Year & Description   &  Backg. & TCI & Metrics & datasets  & MLT & DLT  & Apps & RGCs & FDs \\ \hline

\cite{crowson2020machine} & 2019   & The intersection of \ac{ML} and \acp{CI}   &  \xmark & \xmark & \xmark  & \xmark & \cmark & \xmark & \cmark & \ding{72} & \ding{72} \\

\cite{manero2023improving} & 2022& Explores materials and devices designed for medical care & \xmark & \xmark & \xmark & \ding{72}& \xmark & \xmark & \cmark & \xmark & \xmark \\
 
\cite{wilson2022harnessing} & 2022  & AI in otolaryngology and communication sciences  &  \cmark & \xmark  & \xmark   & \xmark  & \cmark & \xmark & \ding{72} & \xmark & \xmark \\

\cite{d2022tele}  & 2022  & Remote audiology services  &  \xmark & \xmark & \xmark & \xmark & \xmark & \xmark & \cmark& \xmark& \xmark\\

\cite{henry2021noise}  & 2023   & Signal processing in \ac{CI} for noise reduction  &  \cmark & \xmark & \xmark  & \xmark & \cmark & \xmark & \ding{72} & \xmark & \ding{72}\\

Our  &  2024   & Advanced \ac{AI} algorithms for \acp{CI}   &  \cmark  & \cmark  & \cmark  & \cmark  & \cmark  & \cmark  & \cmark & \cmark & \cmark \\

\hline
\end{tabular}
\begin{flushleft}
 Abbreviations: Taxonomy in CI (TCI), ML-based techniques (MLT), DL-based techniques (DLT), Applications (apps), Research gaps and challenges (RGCs) , Future directions (FDs)  
\end{flushleft}
\end{table*}

\subsection{Statistics on investigated papers}

\begin{figure}
    \centering
    \includegraphics[scale=0.06]{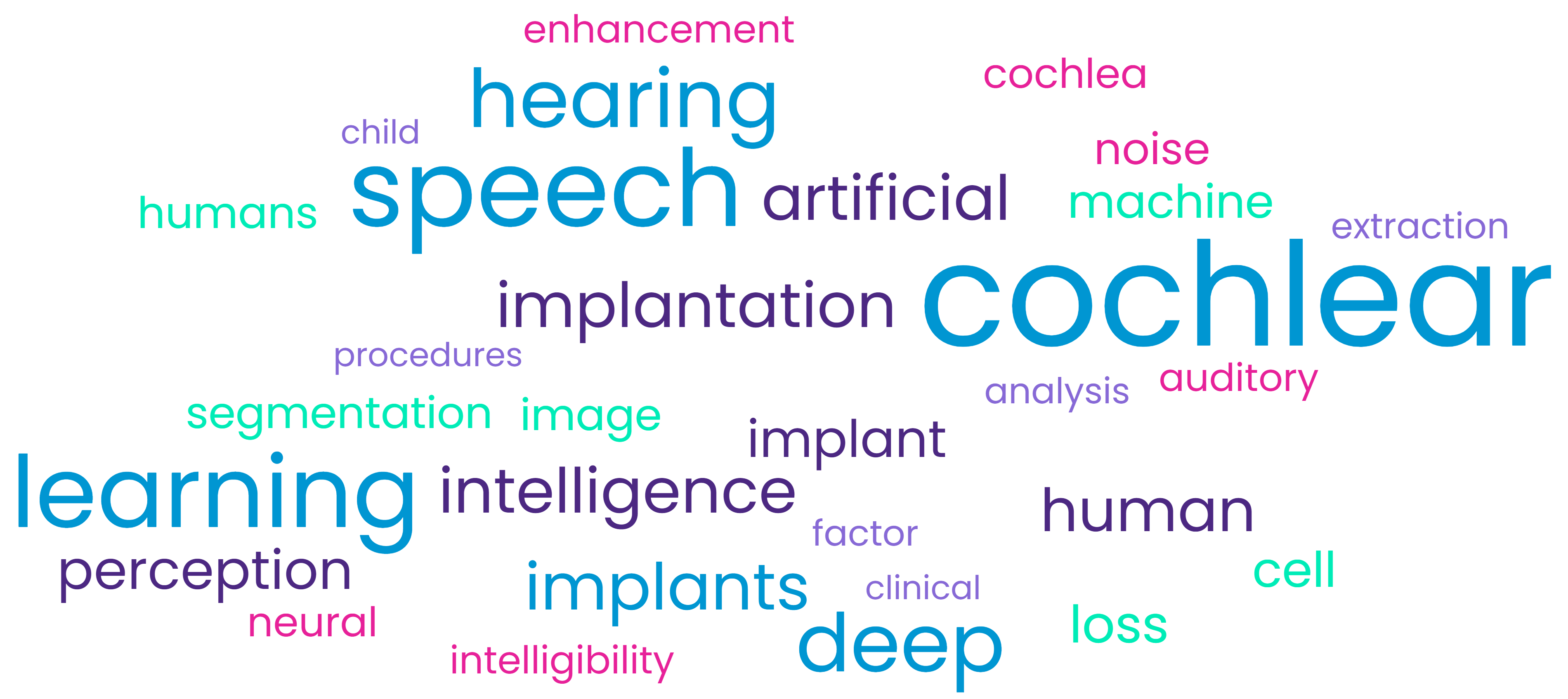}
    \caption{Top keywords in AI-based \ac{CI} research field.}
    \label{fig:keys}
\end{figure}

Recently, there has been a surge in publications related to AI-based \ac{CI}. The review methodology entails defining the search strategy and study selection criteria. Criteria for inclusion, such as keyword relevance and impact, shape the quality assessment protocol. A comprehensive search was conducted on databases such as Scopus and Web of Science. Keywords were extracted for theme clustering, resulting in a formulated query to gather advanced AI-based \ac{CI} studies. The research query retrieves references from papers containing the keywords "Cochlear implant" or "Hearing loss" and "Artificial intelligence" in their abstracts, titles, or authors' keywords. It subsequently refines these papers, focusing on those that also include "Machine learning," "Deep learning," or "Reinforcement learning." Figure \ref{fig:keys} illustrates the most frequently used keywords by the authors in the titles, abstracts, and keywords of the selected papers.

\begin{figure*}
    \centering
    \includegraphics[scale=0.75]{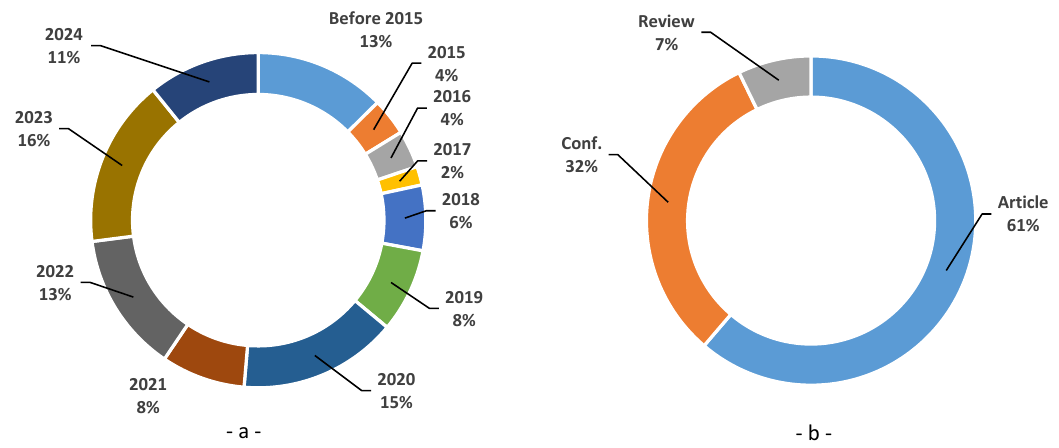}
    \caption{Bibliometrics analysis of the papers included in this review. (a) Papers distribution over the last years. (b) Percentage breakdown of paper types included in this review.}
    \label{fig:bibstat}
\end{figure*}

Figure \ref{fig:bibstat} illustrates the distribution of these papers by type and publication year, encompassing articles, conference papers, and reviews. Additionally, it showcases the percentage of publications from before 2015 to 2023, with a total count of 99 papers. Notably, the peak in publications within this field is observed in 2023 (Figure \ref{fig:bibstat} (a)), underscoring the heightened interest among researchers. Furthermore, the predominant publication type is research articles, followed by conference papers (Figure \ref{fig:bibstat} (b)).

\subsection{Motivation and contribution}

The motivation behind conducting a comprehensive review on \acp{CI} stems from the imperative to critically assess and consolidate the current state of \ac{AI} applications in this crucial field. \acp{CI} have revolutionized auditory rehabilitation for individuals with hearing impairment, and integrating \ac{ML} and \ac{DL} techniques holds immense potential for further advancements. This review seeks to fill a significant gap in the existing literature by providing a detailed analysis of recent \ac{AI}-based \ac{CI} frameworks.

The primary objective is to present a nuanced understanding of the landscape, categorizing frameworks based on \ac{ML} and \ac{DL} methodologies, available datasets, and key metrics. By addressing this gap, the review aims to offer valuable insights for researchers, clinicians, and technologists involved in the development and improvement of \ac{CI} technologies. Furthermore, the exploration of advanced \ac{DL} algorithms, such as transformers and \ac{RL}, in the context of \acp{CI}, underscores the potential for transformative breakthroughs. Ultimately, this research review aspires to contribute to the enhancement of \ac{CI} technologies, fostering innovation and improving the quality of life for individuals with hearing impairment. The principal contributions of this paper can be succinctly outlined as follows:

\begin{itemize}
    \item Detailing the assessment metrics associated with \ac{AI} and \acp{CI}, and elucidating the extensively utilized datasets, whether publicly accessible or generated, employed to validate AI-based \ac{ASR} for \ac{CI} methodologies.
    \item The implementation of \ac{CI}, along with a comprehensive elucidation of the taxonomy encompassing \ac{ML} and \ac{DL}-based \acp{CI}, is thoroughly expounded upon. Additionally, recommended frameworks for AI-based \ac{CI} are thoroughly discussed and succinctly summarized in tables for enhanced clarity.
    \item Providing detailed insights into the applications of \ac{ML} and \ac{DL} within the domain of \ac{CI}, encompassing functions such as denoising and speech enhancement, segmentation, thresholding, imaging, as well as \ac{CI} localization, along with various other functionalities.
    \item Delving into the existing gaps in AI-driven \ac{CI}, offering insights, and proposing novel ideas to address these gaps. Additionally, exploring potential avenues for future research to deepen comprehension and provide valuable guidance for subsequent investigations.
\end{itemize}

The subsequent sections are organized as follows: Section \ref{sec2} delves into the background in speech processing, outlining datasets and metrics. Section \ref{sec3} discusses the methodology employed for \ac{CI} based on \ac{AI}. Section \ref{sec4} presents the medical applications and impact of applying \ac{AI} on \ac{CI}. Section \ref{sec5} offers a comprehensive discussion on research gaps, future directions, and perspectives. Finally, Section \ref{sec6} concludes the paper with implications and future research directions.

%\subsection{Roadmap and paper structure}

%\textcolor{red}{this part is finished when all review is written}

\section{Background in speech processing for \acp{CI} }
\label{sec2}

\subsection{Speech processing for \acp{CI}}
\begin{figure}
    \centering
    \includegraphics[scale=0.5]{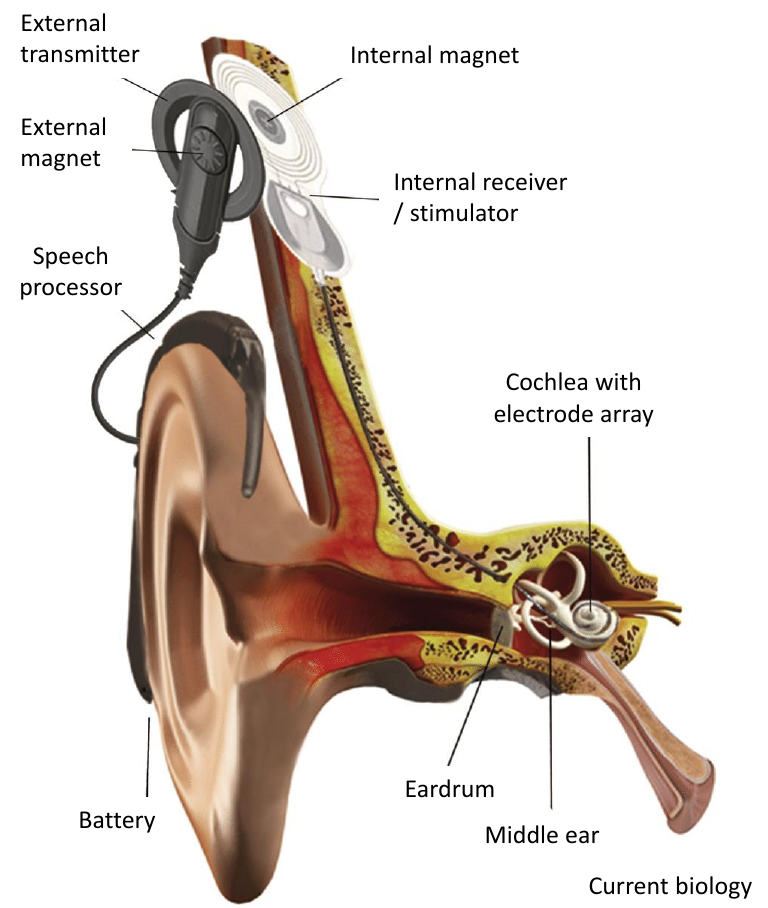}
    \caption{Illustration of a \ac{CI} depicting the components situated externally and internally within the device \cite{macherey2014cochlear}}
    \label{fig:img_coch}
\end{figure}

\acp{CI} are electronic devices that can be implanted in one or both ears to restore some level of hearing for individuals with partial or severe deafness. \acp{CI} comprise an external part with a microphone and speech processor and an internal part with a receiver-stimulator and electrode array, as shown in Figure \ref{fig:img_coch}. They convert sounds into electrical signals, stimulating the auditory nerve to enable sound perception in individuals with profound hearing loss \cite{macherey2014cochlear}.

The incoming sound is divided into multiple frequency channels using bandpass filters and then processed by envelope detectors. Non-linear compressors adjust the dynamic range of the envelope for each patient. The compressed envelope amplitudes are then utilized to modulate a fixed-rate biphasic carrier signal. A current source converts voltage into pulse trains of current, which are delivered to electrodes placed along the cochlea in a non-overlapping manner. This stimulation method is called \ac{CIS}. Another coding strategy, known as \ac{ACE}, uses a greater number of channels and dynamically selects the "n-of-m" bands with the largest envelope amplitudes (prior to compression). Only the corresponding "n" electrodes are stimulated. A popular device widely used for \acp{CI}, such as the cochlear nucleus, typically has 22 channels.

The sound processor, which usually contains a microphone, battery, and other components, can be worn either behind-the-ear (BTE) or off-the-ear (OTE). A headpiece holds a transmitter coil, positioned externally above the ear, while internally, a receiver coil, stimulator, and electrode array are implanted. The SP includes a digital signal processor (DSP) with memory units (maps) that store patient-specific information. An audiologist configures these maps during the fitting process, adjusting thresholds for each electrode, including T-Levels (the softest current levels audible to the \ac{CI} user) and C/M-Levels (current levels perceived as comfortably loud), as well as the stimulation rate or programming strategy. Data (pulse amplitude, pulse duration, pulse gap, etc.) and power are sent through the skull via a radio frequency signal from the transmitter coil to the receiver coil. The stimulator decodes the received bitstream and converts it into electric currents to be delivered to the cochlear electrodes. High-frequency signals stimulate electrodes near the base of the cochlea, while low-frequency signals stimulate electrodes near the apex.

The \ac{CI} stimulates the auditory nerve afferents, which connect to the central auditory pathways. However, compared to individuals with normal hearing (NH), \ac{CI} users face more difficulties in speech perception, particularly in noisy environments. Hearing loss can be caused by various factors, including natural aging, genetic predisposition, exposure to loud sounds, and medical treatments. Damage to the hair cells in the inner ear often leads to a reduced dynamic range of hearing, as well as decreased frequency selectivity and discriminative ability in speech processing. To evaluate the effectiveness of \acp{CI} in speech perception amid noise, listening tests involving both normal hearing individuals and \ac{CI} users are commonly conducted. These tests typically employ a combination of speech utterances from a recognized speech corpus and background noises like speech-weighted noise and babble. Alternatively, vocoder simulations can be utilized alongside speech intelligibility metrics. While sentence-based tests are frequently employed, other stimuli such as vowels, consonants, and phonemes are also used. As a result, noise reduction techniques are increasingly employed to enhance the performance of \acp{CI} in challenging environments \cite{henry2021noise}.

\subsection{Datasets}

Researchers have utilized numerous datasets to validate their proposed schemes, comprising both widely recognized publicly available sets and locally generated ones. These datasets fall into two categories: speech or images. Table \ref{tab:dataset} provides a summary of these datasets, detailing their characteristics, citing studies that have utilized them, and indicating their availability through links or references.

\begin{table*}[h!]
\caption{List of publicly available datasets used for \acp{CI} applications}. 
\label{tab:dataset}
\scriptsize
\begin{tabular}{lm{1cm}m{1cm}m{9.5cm}m{1.5cm}m{0.7cm}}
\hline
Dataset & NoS & NoC & Characteristics & Used by & Source \\
\hline
\acs{MHINT} & 2560 \newline  & 8  & It is a test resource that was developed in two versions: MHINT-M for use in Mainland China and MHINT-T for use in Taiwan. The development of MHINT took into consideration the tonal nature of Mandarin, recognizing the importance of lexical tone in designing the test. & \cite{wang2020improving},\cite{lai2016deep},\cite{wang2017deep} & \cite{wong2007development}\\[0.2cm]

PhonDat 1 & 21587  & 201  & It is from the Bavarian Archive For Speech Signals, contains 21.4 hours of speech data, and it is orthographically transcribed and phonemically annotated. & \cite{arias2021multi},\cite{arias2019multi} & Link\tablefootnote{\url{https://www.phonetik.uni-muenchen.de/Bas/BasPD1eng.html}}\\[0.2cm]

 TIMIT & 6300  & 630 & The dataset consists of phonemically and lexically transcribed speech from American English speakers belonging to diverse demographics and dialects. It provides comprehensive information with time-aligned orthographic, phonetic, and word transcriptions. Additionally, each utterance is accompanied by its corresponding 16-bit, 16kHz speech waveform file, ensuring a complete and detailed dataset for analysis and experimentation in \ac{ASR} and acoustic-phonetic studies. & \cite{hinrichs2022vector},\cite{gajecki2023deep},\cite{chu2018using} & Link\tablefootnote{\url{https://www.kaggle.com/mfekadu/darpa-timit-acousticphonetic-continuous-speech}}\\[0.2cm]

 GSC & 18 Hours & 30 & The dataset was gathered using crowd-sourcing. It consists of 65,000 recordings, each lasting one second, and contains 30 brief words. Among these words, 20 commonly used ones were spoken five times by the majority of participants, while 10 other words (considered unfamiliar) were spoken only once. & \cite{grimm2019simulating} & Link\tablefootnote{\url{https://research.
googleblog.com/2017/08/launching-speech-commands-dataset.html}}\\
 
WSJ0-2mix & - & - & This dataset is a collection of mixed speech recordings that are used for speech recognition. It utilizes utterances from the Wall Street Journal (WSJ0) corpus to create these mixtures. & \cite{feng2022preservation} & Link \tablefootnote{\url{https://www.merl.com/research/highlights/deep-clustering}}\\

LibriVox & 547 Hours & - &  The LibriVox dataset consists of sentence-aligned German audio, text, and English translations from audio books. & \cite{gajecki2023deep} & Link \tablefootnote{\url{https://librivox.org/}}\\

HSM & 20 & 30 & The development of the German HSM Sentence Test aimed to provide an ample amount of test sentences for the repeated assessment of speech comprehension among \ac{CI} users. & \cite{gajecki2023deep},\cite{gajecki2022end} & \cite{hochmair1997hsm}\\

 CEC1 & 6000 & 24 & The dataset features simulated living rooms with static sources, including a single target speaker, interferer (competing talker or noise), and a large target speech database of English sentences produced by 40 British English speakers. & \cite{gajecki2022end} & Link\tablefootnote{\url{https://claritychallenge.org/clarity_CEC1_doc/docs/cec1_data}}\\
 
DEMAND & 560 & 6 & The dataset consists of 15 recordings capturing acoustic noise in various environments. These recordings were made using a 16-channel array, with microphone distances ranging from 5 cm to 21.8 cm. & \cite{gajecki2023deep} & Link\tablefootnote{\url{https://www.kaggle.com/datasets/chrisfilo/demand}}\\
THCHS-30 & 35 Hours & 50 & The dataset is a free Chinese speech corpus accompanied by resources such as lexicon and language models. & \cite{kang2021deep} & Link\tablefootnote{\url{https://www.openslr.org/18/}}\\

BCP & 55938  & 20 & The \ac{BCP} dataset contains Cocktail Party scenarios with individuals wearing \ac{CI} audio processors and a head and torso simulator. Recorded in an acoustic chamber, it includes multi-channel audio, image recordings, and digitized microphone positions for each participant. & \cite{fischer2021speech} & \cite{fischer2020multichannel}\\

ikala & 252-30s & 206 & The dataset comprises audio recordings consisting of vocal and backing track music with a sampling rate of 44100 Hz. Each music track is a stereo recording, where one channel contains the singing voice and the other channel contains the background music. All tracks were performed by professional musicians and featured a group of six singers, evenly split between three females and three males. & \cite{tahmasebi2020design} & Link\tablefootnote{\url{https://zenodo.org/records/3532214}}\\

 MUSDB & 150 & 4 & The dataset is a collection of music tracks specifically designed for music source separation research. It consists of professionally mixed songs across various genres, with individual tracks isolated for vocals, drums, bass, and other accompaniment. & \cite{tahmasebi2020design} & Link\tablefootnote{\url{https://sigsep.github.io/datasets/musdb.html}}\\
 
CQ500 & 491 & - & The dataset includes anonymized dicoms files, along with the interpretations provided by radiologists. The interpretations were conducted by three radiologists who have 8, 12, and 20 years of experience in interpreting cranial CT scans, respectively.& \cite{radutoiu2022accurate} & Link\tablefootnote{\url{http://headctstudy.qure.ai/dataset}}\\

ECochG & 21 & - & The dataset captures inner ear potentials in response to acoustic stimulation, specifically focusing on cochlear implant recipients. & \cite{schuerch2022objectification},\cite{schuerch2023objective},\cite{schuerch2023intracochlear} & Link\tablefootnote{\url{https://zenodo.org/records/7092661}}\\

{LibriSpeech}  & {1000 Hours}  & {2484 speakers} & {  English speech recordings with sampling rate 16kHz, the recordings are sourced from audiobooks read aloud for the LibriVox dataset and have been meticulously segmented and synchronized.} & \cite{borjigin2024deep} & Link\tablefootnote{\url{https://www.openslr.org/12}}\\[0.2cm]

{WHAM!}  & {81.68 Hours}  & {28000 files} & {It contains pairs two-speaker mixtures with unique noise backgrounds. WHAMR! extends this by adding artificial reverberation. Noise audio, collected from various urban locations, is provided with scripts to build the datasets.} & \cite{borjigin2024deep} & Link\tablefootnote{\url{https://wham.whisper.ai/}}\\[0.2cm]

\hline
\end{tabular}
\begin{flushleft}
Abbreviations: Number of samples (NoS), Number of classes (NoC)    
\end{flushleft}
\end{table*}

\textcolor{magenta}{}

\subsection{Metrics}
\label{sec34}

Multiple evaluation metrics are utilized during the training and validation of any \ac{DL} models, including AI-based \ac{CI}. These metrics, integral to the confusion matrix, are widely known and applicable across various data types such as speech or images. They include \ac{Acc}, \ac{Sen}, \ac{Rec}, \ac{Spe}, \ac{Pre}, \ac{F1}, and \ac{ROC}. Moreover, different metrics play roles in prediction tasks. For instance, \ac{IoU} assesses overlap, and \ac{MAE} quantifies absolute differences. For a comprehensive understanding of the metrics discussed, including their equations, refer to the details provided in \cite{kheddar2023deep,habchi2023ai}. Other metrics that are widely used for \ac{CI} are summarized in Table \ref{tab:metrics}.

\begin{table*}[ht!]
\caption{An overview of the metrics employed for evaluating \ac{CI} methods.}
\scriptsize
\begin{tabular}[!t]{m{15mm}m{70mm}m{85mm}}

\label{tab:metrics} \\
\hline
Metric & Formula & Description   \\ 
\hline
\acs{DCS} &\(\displaystyle
 \frac{{2 \cdot |A \cap B|}}{{|A| + |B|}}\)& The metric \ac{DCS} is used to evaluate the performance of the vestibule segmentation network. The Dice coefficient is a widely used similarity metric in image segmentation tasks. It measures the overlap between the predicted segmentation mask and the ground truth mask.\\[0.6cm]
 
\acs{ASD}& \(\displaystyle
    \frac{\sum_{a \in S(A)}{\min_{b \in S(B)}(a - b)} + \sum_{b \in S(B)}{\min_{a \in S(A)}(b - a)}}{|S(A)| + |S(B)|} \) &   \Ac{ASD} is a commonly used evaluation measure in medical image segmentation tasks. It quantifies the average distance between the surfaces of two segmented objects, typically the predicted segmentation and the ground truth.\\ [0.6cm]
\acs{AVD}& \(\displaystyle \max(\text{d}(A, B), \text{d}(B, A)) \) & \Ac{AVD} quantifies the average difference in volume between a predicted segmentation and a reference or ground truth segmentation..\\ [0.4cm]

\acs{SNR}& \(\displaystyle \mathrm{10 \log_{10} \Big(\frac{Signal}{Noise}\Big)_{dB}} \)  & \Ac{SNR} is a measure of the quality of the speech signal. It is commonly used to evaluate the quality of the stego-speech (the speech signal after the hidden information has been embedded). A lower SNR indicates that the steganography technique has introduced more distortion to the speech signal.\\ [4mm]

\acs{ASE} & \(\displaystyle  \frac{\sum E_{ij}}{N} \) &  
 Is the \ac{ASE}.  The distances between corresponding points on the measured surface and the reference surface are computed using \(\displaystyle E_{ij} = |I_1(i, j) - I_2(k, l)| \), known as \ac{P2PE}. $N$ is the total number of correspondence points. Normalize average error by dividing by the imaging system's dynamic range. $(i,j)$, and  $(k,l)$ represent pixel values in the first and second images, respectively.\\[4mm]

\acs{ILD} &   \( 20 \log_{10} \left( \frac{L}{R} \right) \) &  \Ac{ILD} a psychoacoustic metric, measuring sound level differences between left (L) and right (R) ears, reflects cues essential for sound localization, enhancing spatial awareness in auditory perception for directional sound source identification.\\[5mm]

\acs{MEE}& \(\displaystyle   \frac{1}{N} \sum_{i=1}^{N} |L_i - \hat{L}_i|  \)   & \Ac{MEE} measures the average absolute difference between the true \(L_i\) and estimated \(\hat{L}_i\) endpoint locations across multiple utterances \(N\).   \\

\acs{MMSE} &  Estimator \(\displaystyle \hat{x}_{MMSE} = E[ X | Y] \)  & \Ac{MMSE}  is a statistical estimation technique used in speech enhancement to minimize the mean square error between the estimated  \(Y\) and true \(X\) clean speech signals \\

\acs{STOI} & \(\displaystyle  \frac{\sum_{t=1}^{T} g(t) \cdot \text{STOI}_{\text{frame}}(t)}{\sum_{t=1}^{T} g(t)}  \)   &  \Ac{STOI} is a metric used to assess the intelligibility of time-frequency weighted noisy speech. It is based on the idea that human speech perception relies on the availability of important acoustic features in short time frames \cite{taal2010short}.  \\

\acs{NCM} &   & \Ac{NCM}\\

\acs{SDR}, \acs{SAR}, and \acs{SIR} & \(\displaystyle 10log_{10}\frac{{\left\|s_{target}  \right\|}^{2}}{{\left\|e_{total} \right\|}^{2}}, \hspace{0.3cm} 10log_{10}\frac{{\left\|s_{target}+e_{inter}+e_{noise}  \right\|}^{2}}{{\left\|e_{artifacts}  \right\|}^{2}},\newline \hspace{0.3cm} 10log_{10}\frac{{\left\|s_{target}  \right\|}^{2}}{{\left\|e_{inter}  \right\|}^{2}} \) &  \Ac{SDR}, \ac{SAR}, and \ac{SIR} are metrics objectively assess and compare speech source-separation algorithms based on accuracy and minimization of distortions and interference. \ac{SDR} gauges source separation quality by comparing true source power to introduced distortion. \ac{SAR} evaluates source separation from artifacts or noise, while \ac{SIR} measures the ratio of true source power to interference after separation. \\
 \hline
 \end{tabular}
\end{table*}
%\end{center}

\section{Taxonomy of \ac{CI}-based AI techniques }
\label{sec3}

\begin{figure*}[h!]
    \centering
    \includegraphics[scale=0.8]{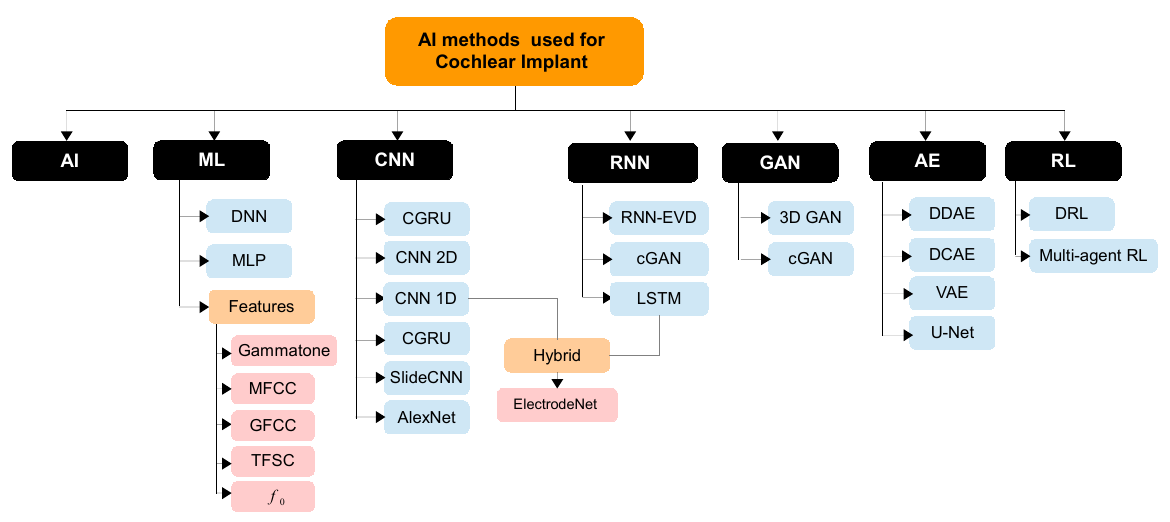}
    \caption{Taxonomy of the employed AI techniques for \ac{CI}.}
    \label{fig:taxAI}
\end{figure*}

Several artificial intelligence techniques have been employed to enhance the efficacy of \acp{CI}. While some rely on 1D data, others process information in a 2D image format. Figure \ref{fig:taxAI} summarizes all AI algorithms utilized, alongside the features employed and hybrid AI methodologies. Additionally, Table \ref{tab} provides a summary of DL-based techniques utilized in \ac{CI} hearing devices.

\subsection{\ac{CI}-based AI Implementation}
\label{sec23}

\begin{figure}[ht!]
    \centering
    \includegraphics[scale=0.33]{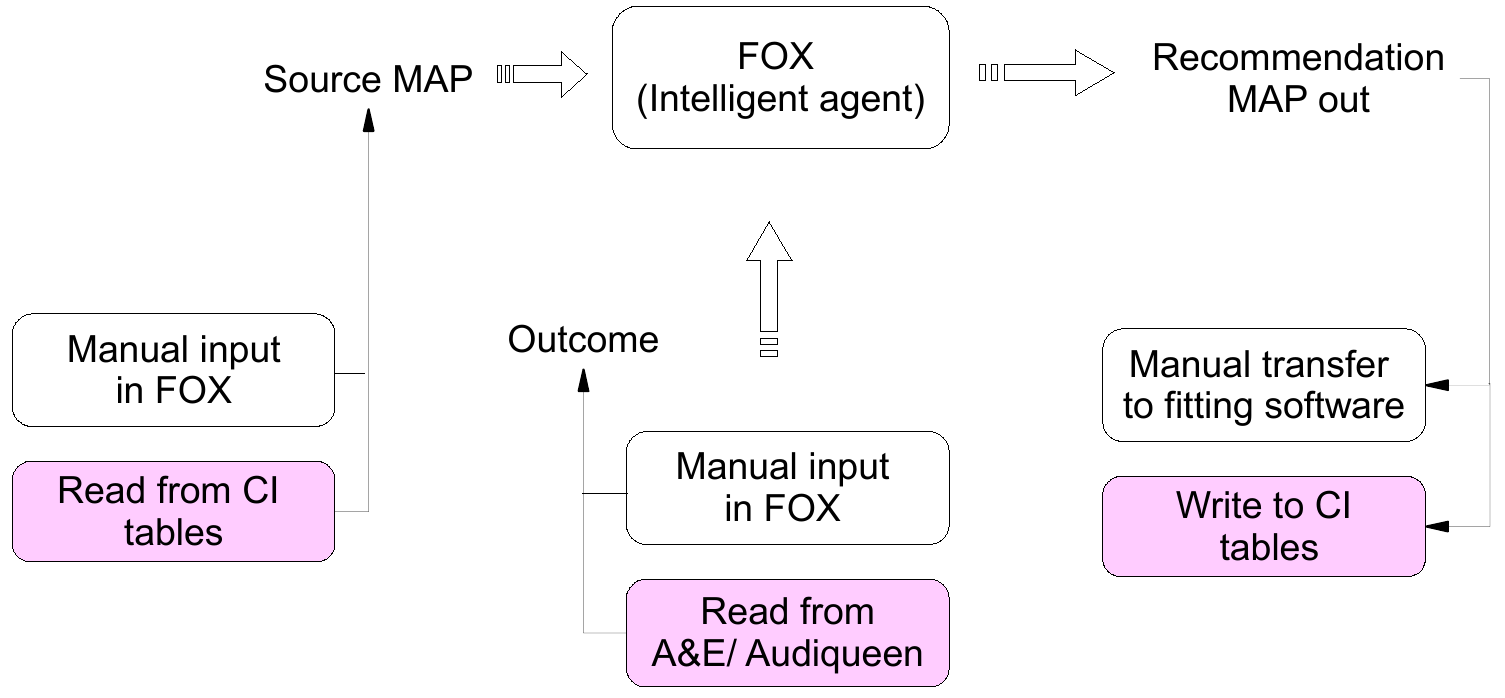}
    \caption{ The fundamental operating concept of the \ac{FOX} involves inputting an initial program and multiple psychoacoustic test outcomes. \ac{FOX} processes this information and generates fitting suggestions as its output. When integrated with proprietary outcome and \ac{CI} fitting software, the shaded boxes represent its functionality, while the unfilled boxes represent its standalone capability \cite{govaerts2010development}. Audiqueen is a dataset with A and E (A\&E) phoneme discrimination.} 
    \label{fig:prog1}
\end{figure}

\Ac{CI} programming involves adjusting device settings to optimize sound perception for individual users. This includes setting stimulation levels, electrode configurations, and signal processing parameters to enhance speech understanding and auditory experiences based on patient feedback and objective measures. In 2010, Govaerts et al. \cite{govaerts2010development} described the development of an intelligent agent, called \ac{FOX}, for optimizing \ac{CI} programming, as illustrated in Figure \ref{fig:prog1}. The agent analyzes map settings and psychoacoustic test results to recommend and execute modifications to improve outcomes. The tool focuses on an outcome-driven approach, reducing fitting time and improving the quality of fitting. It introduces principles of \ac{AI} into the \ac{CI} fitting process. The study proposed objective measures and group electrode settings as strategies to reduce fitting time.

Similarly, in \cite{vaerenberg2011experiences,meeuws2017computer,wathour2020manual,waltzman2020use,wathour2023prospective}, all have employed \ac{FOX} for programming \ac{CI}. Vaerenberg et al. \cite{vaerenberg2011experiences} discusses the use of \ac{FOX} for programming \ac{CI} sound processors in new users. \ac{FOX} modifies maps based on specific outcome measures using heuristic logic and deterministic rules. The study showed positive results and optimized performance after three months of programming, with good speech audiometry and loudness scaling outcomes. The paper highlights the importance of individualized programming parameters and the need for outcome-based adjustments rather than relying solely on comfort. In \cite{meeuws2017computer}, computer-assisted \ac{CI} fitting using \ac{FOX} assessed its impact on speech understanding. Results from 25 recipients showed that 84\% benefited from suggested map changes, significantly improving speech understanding thanks to the learning capacity of \ac{FOX}. This approach offers standardized, systematic \ac{CI} fitting, enhancing auditory performance.

The COCH gene, referred to as the Cochlin gene, is responsible for encoding the cochlin protein situated on chromosome 14 in humans, primarily expressed in the inner ear. Cochlin predominantly functions within the cochlea, a spiral-shaped structure involved in the process of hearing, contributing to its structural integrity and proper operation. Wathour et al. in \cite{wathour2020manual} discuss the use of \ac{AI} in \ac{CI} fitting through two case studies. The first case involves a 75-year-old lady who received a left ear implant due to gradual and severe hearing loss in both ears without a clear cause. In the second case, a 72-year-old man with a COCH gene mutation causing profound hearing loss in both ears underwent a right ear implant to assess whether \ac{CI} programming using the \ac{AI} software \ac{FOX} application could improve \ac{CI} performance. The results showed that \ac{AI}-assisted fitting led to improvements in auditory outcomes for adult \ac{CI} recipients who had previously undergone manual fitting. The \ac{AI} suggestions helped improve word recognition scores and loudness scaling curves. Similarly, Waltzman et al. \cite{waltzman2020use} incorporate \ac{AI} in programming \acp{CI}, aiming to assess the performance and standardization of \ac{AI}-based programming on fifty-five adult \ac{CI} recipients. The results showed that the \ac{AI}-based \ac{FOX} system performed better for some patients, while others had similar results; however, the majority preferred the \ac{FOX} system.

\subsection{\Ac{ML}-based methods}

\ac{ML} is a subfield of \ac{AI} that focuses on developing algorithms and statistical models that enable computer systems to improve their performance in a specific task by learning features from input data. The research in \cite{eichler2022algorithm} focuses on algorithm-based hearing and speech therapy rehabilitation after cochlear implantation, particularly for older individuals. They propose the development of an \ac{ML}-based application that offers personalized hearing therapy tailored to the patient's needs, such as select exercises, adjust difficulty levels, and analyze patient difficulties. It operates independently, reducing the reliance on local speech therapists, cost-effective, and accessible alternative to traditional therapy, improving outcomes and quality of life for \ac{CI} recipients. 
In addition, Torresen et al. \cite{torresen2016data} discusses the use of \ac{ML} techniques to streamline the adjustment process for \acp{CI}. The goal is to predict optimal adjustment values for new patients based on data from previous patients. By analyzing data from 158 former patients, the study shows that while fully automatic adjustments are not possible, \ac{ML} can provide a good starting point for manual adjustment. The research also identifies the most important electrodes to measure for predicting levels of other electrodes. This approach has the potential to reduce programming time, benefit patients, and improve speech recognition scores, particularly for young children and patients with post-lingual deafness. Henry et al. in their  \cite{henry2023experimental} investigates the importance of acoustic features in optimizing intelligibility for \acp{CI} in noisy environments. The study employs \ac{ML} algorithms and extracts acoustic features from speech and noise mixtures to train a \ac{DNN}. The results, using various metrics, reveal that frequency domain features, particularly Gammatone features, perform best for normal hearing, while Mel spectrogram features exhibit the best overall performance for hearing impairment. The study suggests a stronger correlation between \ac{STOI} and \ac{NCM} in predicting intelligibility for hearing-impaired listeners. The findings can aid in designing adaptive intelligibility enhancement systems for \acp{CI} based on noise characteristics.

\begin{figure}[ht!]
    \centering
    \includegraphics[scale=0.85]{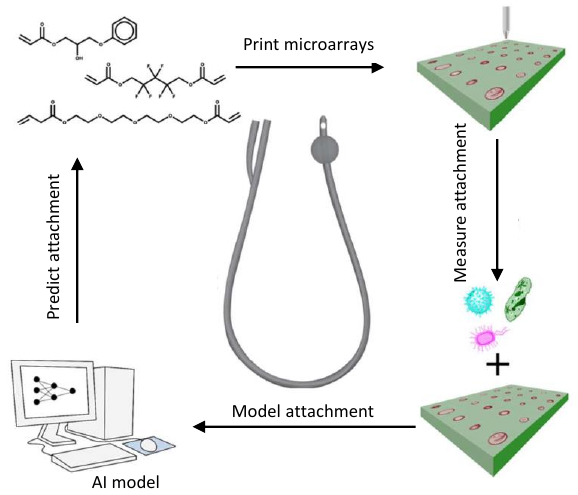}
    \caption{ Diagram illustrating the procedures utilized in the production of microarrays, analysis of pathogens, data modeling, and forecasting the attachment of pathogens to novel polymers  \cite{mikulskis2018prediction}.}
    \label{fig:ML1}
\end{figure}

Moreover, the research in \cite{pavelchek2023imputation} focuses on imputing missing values in \ac{CI} candidate audiometric data. This study assessed the performance of various imputation algorithms using a dataset of 7,451 audiograms from \ac{CI} patients. The results showed that the quantity of missing data affected the imputation performance, with greater amounts leading to poorer results. The distribution of sparsity in the audiometric data was found to be non-uniform, with inter-octave frequencies being less commonly tested. The \ac{MICE} method, safely imputed up to six missing data points in an 11-frequency audiogram, consistently outperformed other models. This study highlights the importance of imputation techniques in maximizing datasets in hearing healthcare research. Xu et al. in \cite{xu2023objective} explores the objective discrimination of bimodal speech using \acp{FFR}. The study investigates the neural encoding of \ac{$f_0$}, called also pitch \cite{kheddar2019pitch},  and \ac{TFSC} in simulated bimodal speech conditions. The results show that increasing acoustic bandwidth enhances the neural representation of \ac{$f_0$} and \ac{TFSC} components in the non-implanted ear. Moreover, \ac{ML} algorithms successfully classify and discriminate \acp{FFR} based on spectral differences between vowels. The findings suggest that the enhancement of \ac{$f_0$} and \ac{TFSC} neural encoding with increasing bandwidth is predictive of perceptual bimodal benefit in speech-in-noise tasks. \acp{FFR} may serve as a useful tool for objectively assessing individual variability in bimodal hearing. The research conducted by Crowson et al.  \cite{crowson2020predicting} aimed to predict postoperative \ac{CI} performance using supervised \ac{ML}. The authors used neural networks and \ac{DT}-based ensemble algorithms on a dataset of 1,604 adults who received \acp{CI}. They included 282 text and numerical variables related to demographics, audiometric data, and patient-reported outcomes. The results showed that the neural network model achieved a 1-year postoperative performance prediction \ac{RMSE} of 0.57 and classification accuracy of 95.4\%. When both text and numerical variables were used, the \ac{RMSE} was 25.0\% and classification accuracy was 73.3\%. The study identified influential variables such as preoperative sentence-test performance, age at surgery, and specific questionnaire responses. The findings suggest that supervised \ac{ML} can predict \ac{CI} performance and provide insights into factors affecting outcomes.
In the same context of prediction, Mikulskis et al. \cite{mikulskis2018prediction} focuse on predicting the attachment of broad-spectrum pathogens to coating materials for biomedical devices as illustrated in Figure \ref{fig:ML1}. The authors employ \ac{ML} methods to generate quantitative predictions for pathogen attachment to a large library of polymers. This approach aims to accelerate the discovery of materials that resist bacterial biofilm formation, reducing the rate of infections associated with medical devices. The study highlights the need for new materials that prevent bacterial colonization and biofilm development, particularly in the context of antibiotic resistance. The results demonstrate the potential of \ac{ML} in designing polymers with low pathogen attachment, offering promising candidate materials for implantable and indwelling medical devices. Similarly, Alohali et al. \cite{alohali2023machine} focuses on using \ac{ML} algorithms to predict the post-operative electrode impedances in \ac{CI} patients. The study used a dataset of 80 pediatric patients and considered factors such as patient age and intraoperative electrode impedance. The results showed that the best algorithm varied by channel, with Bayesian linear regression and neural networks providing the best results for 75\% of the channels. The accuracy level ranged between 83\% and 100\% in half of the channels one year after surgery. Additionally, the patient's age alone showed good prediction results for 50\% of the channels at six months or one year after surgery, suggesting it could be a predictor of electrode impedance.

{Recently, Zeitler et al. \cite{zeitler2024predicting} developed supervised \ac{ML} classifiers to predict acoustic hearing preservation in patients undergoing \ac{CI} surgery. The classifiers were trained using preoperative clinical data from 175 patients. The analysis revealed associations between various factors and hearing preservation outcomes. The random forest classifier demonstrated the highest mean performance in predicting outcomes. \ac{ML} showed potential for predicting residual acoustic hearing preservation and improving clinical decision-making in cochlear implantation.}

\subsection{\acs{CNN}-based methods}

\Acp{CNN} are a class of \ac{DL} algorithms widely used in computer vision tasks. Their architecture includes convolutional layers that automatically learn hierarchical features from input data. The core convolution for 2D data operation is defined by the equation:

\begin{equation}
  S(i, j) = (I * K)(i, j) = \sum_m \sum_n I(m, n) \cdot K(i - m, j - n)  
\end{equation}

Here, \(I\) represents the input 2D data, \(K\) is the convolutional kernel, and \(S\) is the output feature map. \Acp{CNN} excel at recognizing spatial patterns, making them essential in image recognition, object detection, and other visual tasks. Additionally, there exist 1D \acp{CNN}, which are effective for sequential data analysis, such as in natural language processing or time series applications. \ac{CNN} is widely used for the interdisciplinary nature of \ac{CI}, which involves aspects of neurobiology, signal processing, and medical technology.
For example, the proposed work \cite{islam2022novel} introduces a novel pathological voice identification system using signal processing and \ac{DL}. It employs \ac{CI} models with bandpass and optimized gammatone filters to mimic human cochlear vibration patterns. The system processes speech samples and utilizes a \Ac{CNN} for final pathological voice identification. Results show discrimination of pathological voices with \ac{F1} scores of 77.6\% (bandpass) and 78.7\% (gammatone). The paper addresses voice pathology causes, compares filter models, and proposes a non-invasive, objective assessment system. It contributes to the field with a comprehensive performance analysis, achieving high accuracy and demonstrating effectiveness compared to related works. Addtionally, in the scheme proposed by Wang \cite{wang2020improving}, the \ac{FCN} model is evaluated for enhancing speech intelligibility in mismatched training and testing conditions. Using 2,560 Mandarin utterances and 100 noise types, the study compares \ac{FCN} with traditional \ac{MMSE} and \ac{DDAE} models. Two sets of experiments are conducted for normal and vocoded speech. The \ac{FCN} model demonstrates superior performance, maintaining clearer speech structures, especially in mid-low frequency regions crucial for intelligibility. Objective evaluations using \ac{STOI} scores and a listening test confirm \ac{FCN}'s effectiveness under challenging \ac{SNR} conditions, outperforming \ac{MMSE} and \ac{DDAE}. The study suggests \ac{FCN} as a promising choice for \ac{EAS} speech processors. 

Moving on, the research paper in \cite{de2023optimizing}  presents a novel approach to optimize stimulus energy for \acp{CI}. A \ac{CNN} was developed as a surrogate model for a biophysical auditory nerve fiber model, significantly reducing simulation time while maintaining high accuracy. The \ac{CNN} was then used in conjunction with an evolutionary algorithm \cite{back1993overview} to optimize the shape of the stimulus waveform, resulting in energy-efficient waveforms  and potential improvements in \ac{CI} technology.  Traditional computational models of the cochlea, which represent it as a transmission line, are computationally expensive due to their cascaded architecture and the inclusion of non-linearities. As a result, they are not suitable for real-time applications such as hearing aids, robotics, and \ac{ASR}. For the aforementioned conditions, the study in \cite{baby2021convolutional} presents a hybrid approach, called CoNNear\footnote{\url{https://github.com/HearingTechnology/CoNNear_periphery}}, which combines \acp{CNN}, capable of performing end-to-end waveform predictions in real-time, with computational neuroscience to create a real-time model of human cochlear mechanics and filter tuning. The \ac{CNN} filter weights were trained using simulated \ac{BM} displacements from cochlear channels, and the model's performance was evaluated using basic acoustic stimuli. The CoNNear model is designed to capture the tuning, level-dependence, and longitudinal coupling characteristics of human cochlear processing. It converts acoustic speech stimuli into \ac{BM} displacement waveforms across 201 cochlear filters. Its computational efficiency and ability to capture human cochlear characteristics make it suitable for developing human-like machine-hearing applications.

The research paper in \cite{grimm2019simulating} explores the utilization of a  \ac{CNN} in simulating speech processing with \acp{CI}. The study investigates the effect of channel interaction, a phenomenon that degrades spectral resolution in \ac{CI} delivered speech, on learning in neural networks. By modifying speech spectrograms to approximate \ac{CI} delivered signals, the \ac{CNN} is trained to classify them. The findings suggest that early in training, the presence of channel interaction negatively impacts performance. This indicates that the spectral degradation caused by channel interaction conflicts with perceptual expectations acquired from high-resolution speech. The study highlights the potential for reducing channel interaction to enhance learning and improve speech processing in \ac{CI} users, particularly those who have adapted to high-resolution speech.

Schuerch et al. \cite{schuerch2022objectification} focus on the objectification of \ac{ECochG} using AlexNet, \ac{CNN} architecture, to automate and standardize the assessment and analysis of \ac{CM} signals in \ac{ECochG} recordings for clinical practice and research. The authors compared three different methods: correlation analysis, Hotelling's T2 test, and \ac{DL}, to detect \ac{CM} signals. The \ac{DL} algorithm performed the best, followed closely by Hotelling's T2 test, while the correlation method slightly underperformed. The automated methods achieved excellent discrimination performance in detecting \ac{CM} signals with an accuracy up to 92\%, providing fast, accurate, and examiner-independent evaluation of \ac{ECochG} measurements.

\begin{figure*}
    \centering
    \includegraphics[scale=1.1]{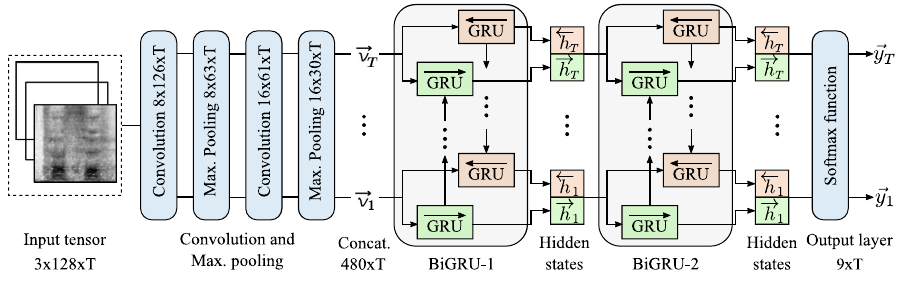}
    \caption{ \ac{CGRU} architecture, focusing on input sequences composed of 3D-channel data generated from Mel-spectrograms, Cochleagrams, and \ac{CWT} using Morlet wavelets \cite{arias2021multi}.}
    \label{fig:cnn1}
\end{figure*}

Moreover, Arias et al. \cite{arias2021multi} presents a methodology for speech processing using \acp{CNN}. The study aims to improve the representation learning capabilities of \acp{CNN} by combining multiple time-frequency representations of speech signals. The proposed approach involves generating multi-channel spectrograms by combining continuous wavelet transform, Mel-spectrograms, and Gammatone spectrograms. These spectrograms are utilized as input data for the \ac{CNN} models. The effectiveness of the approach is evaluated in two applications: automatic detection of speech deficits in \ac{CI} users and phoneme class recognition. The results demonstrate the advantages of using multi-channel spectrograms with \acp{CNN}, showcasing improved performance in speech analysis tasks. The \ac{CGRU} architecture is utilized,  as illustrated in Figure \ref{fig:cnn1}. The input sequences consist of 3D-channel inputs created by combining Mel-spectrograms, Cochleagrams, and \ac{CWT} with Morlet wavelets. Convolution is applied solely on the frequency axis in order to preserve the time information. The resulting feature maps are subsequently fed into a 2-stacked bidirectional \ac{GRU}. A softmax function is employed to predict the phoneme label for each speech segment in the input signal.

This paper \cite{arias2019multi} introduces a novel method for automatically detecting speech disorders in \ac{CI} users using a multi-channel \ac{CNN}. The model processes 2-channel input comprising Mel-scaled and Gammatone filter bank spectrograms derived from speech signals. Testing on 107 \ac{CI} users and 94 healthy controls demonstrates improved performance with 2-channel spectrograms. The study addresses a gap in acoustic analysis of \ac{CI} user speech, proposing a \ac{DL} approach with potential applications beyond \ac{CI} users. Experimental results indicate the effectiveness of the proposed \ac{CNN}-based method, offering promise for speech disorder detection and potential extensions to other pathologies or paralinguistic aspects that employ \acp{MFCC} and \acp{GFCC} features.

%\cite{schuerch2023intracochlear}

For 2D \ac{CNN}, the following work \cite{zhang2018automatic} introduces \ac{IGCIP}, enhancing \ac{CI} outcomes using image processing. \ac{IGCIP} segments intra-cochlear anatomy in \ac{CT} images, aiding electrode localization for programming. The scheme addresses challenges in automating this process due to varied image acquisition protocols. The proposed solution employs a \ac{DL}-based approach, utilizing \acp{CNN} to detect the presence and location of inner ears in head \ac{CT} volumes. The \acp{CNN} is trained on a dataset with 95.97\% classification accuracy. Results indicate potential for automatic labeling of \ac{CT} images, with a focus on further 3D algorithm development. However, in \cite{de2023optimizing} presents a machine-learning approach to optimize stimulus energy for \acp{CI}. A \ac{CNN} was developed as a surrogate model for a biophysical auditory nerve fiber model, significantly reducing simulation time while maintaining high accuracy. The \ac{CNN} was then used in conjunction with an evolutionary algorithm to optimize the shape of the stimulus waveform, resulting in energy-efficient waveforms. The proposed surrogate model offers an efficient replacement for the original model, allowing for larger-scale experiments and potential improvements in \ac{CI} technology.

The work proposed by \cite{li2022poster}  introduces sliding window based \ac{CNN} (SlideCNN), a novel \ac{DL} approach for auditory spatial scene recognition with limited annotated data. The proposed method converts auditory spatial scenes into spectrogram images and utilizes a SlideCNN for image classification. Compared to existing models, SlideCNN achieves a significant improvement in prediction accuracy, with a 12\% increase. By leveraging limited annotated samples, SlideCNN demonstrates an 85\% accuracy in detecting real-life indoor and outdoor scenes. The results have practical implications for analyzing auditory scenes with limited annotated data, benefiting individuals with hearing aids and \acp{CI}.

This paper \cite{laves2019deep} focuses on advancing laser bone ablation in microsurgery using 4D \ac{OCT}. The challenge lies in automatic control without external tracking systems. The paper introduces a 2.5D scene flow estimation method using \ac{CNN} for \ac{OCT} images, enhancing laser ablation control. A two-stage approach involves lateral scene flow computation followed by depth flow estimation. Training is semi-supervised, combining ground truth error and reconstruction error. The method achieves a \ac{MEE} of (4.7 $\pm$ 3.5) voxel, enabling markerless tracking for image guidance and automated laser ablation control in minimally invasive cochlear implantation. 
{Recently, Almansi et al. \cite{almansi2024novel} presents a radiological software prototype for detecting and classifying normal and malformed inner ear anatomy using cropping algorithms and \ac{CNN} to analyze \ac{CT} images. The software achieved an average accuracy of 92.25\% for cropping inner ear volumes and an AUC of 0.86 for classifying normal and abnormal anatomy. Additionally, Jehn et al. \cite{jehn2024cnns} aimed to improve \ac{AAD} for \ac{CI} users using a \ac{CNN}. EEG data from 25 \ac{CI} users showed that the \ac{CNN} decoder achieved a maximum decoding accuracy of 74\% for a decision window of 60 seconds. Besides, the work in \cite{islam2024cochleagram} introduces a method for detecting dysphonic voice using cochleagram images and a pre-trained \ac{CNN}, achieving 95\% accuracy with sentence samples. However, \cite{cai2024clinical} proposes a method combining high-resolution spiral \ac{CT} scanning with \ac{DL} technique for diagnosing auriculotemporal and ossicle-related diseases. The study utilizes CNN-UNet model to extract sub-pixel information from medical photos of the cochlea. The results demonstrate that this approach improves diagnostic efficiency and enhances understanding of these complex diseases.}

\begin{table*}[h!]
\caption{Summary of some proposed methods based on different \ac{DL} techniques. When comparing the work with numerous existing schemes, only the best-performing one will be highlighted. }
\label{tab:AI-tech}
\scriptsize
\resizebox{\textwidth}{!}{ 
\begin{tabular}{m{0.5cm}m{1cm}m{1cm}m{1.1cm}m{6cm}m{1.5cm}m{2cm}m{1.5cm}}
\hline
Ref. & DL  method & Domain & Dataset & Explanation & CTODLM & Best performance & Improvement  \\
\hline
\cite{wang2020improving} &  \ac{FCN} & Speech & MHINT & Experiment the effectiveness of \ac{FCN} in restoring clean speech signals from noisy counterparts and enhancing the intelligibility of speech in \ac{EAS} devices. &  \ac{DDAE} & \ac{STOI}=0.85 (\ac{SNR}=9dB) &  \ac{STOI}=0.03  \\ [0.2cm]

\cite{arias2021multi} &  \ac{CGRU} & Speech &  PhonDat 1 & Using of onset and offset  
transitions in speech analysis 
and  The input 
sequences are 3D-channel inputs  & \ac{CGRU}  & \ac{Pre}=0.68 
 $\pm\ 0.22$ \newline \ac{Rec}=0.69 $\pm\ 0.21$ \newline \ac{F1}=0.69 $\pm\ 0.19$ &  
 ---  \\ [0.2cm]
 
\cite{laves2019deep} &  \ac{CNN} & Image & Created  & Multiple stages and follows a pyramidal network architecture inspired  by SPyNet, which is designed for optical flow estimation.  & ---  & 
     \ac{MEE}=4.7$\pm\ 3.5$ &  ---  \\ [0.2cm]
     
\cite{arias2019multi} &  multi-channel \ac{CNN} & Speech & PhonDat 1 & Developing a methodology for the automatic detection of speech disorders in \ac{CI} users using a 2D-channel \ac{CNN}. & SVM & \ac{Acc}=83.4,\newline \ac{Sen}=91.7,\newline \ac{Spe}=78.7  &  \ac{Acc}=+0.6,  
   \newline \ac{Sen}=-1.8, \newline \ac{Spe}=+2.7  \\ [0.2cm]
   
\cite{lai2016deep} & DDAE  & Speech & MHINT & The architecture of the DDAE model consisting of five layers with 500 neurons each. Objective evaluations were conducted using STOI and NCM, and a listening test (\acs{PESQ}) &  \acs{KLT}  & \ac{NCM}=0.25 (\ac{SNR}=9dB ) \newline \ac{STOI}=0.85 (\ac{SNR}=12dB) &  \ac{STOI}=+0.07 \\ [0.2cm]  

\cite{chen2023deeply} & Res-CA  & Image &  Created & Introduces Res-CA block, GCPFE, ACE loss, and DS mechanism to enhance the DAE's performance and robustness for compressing CI stimulation patterns. & 3D-DSD & \ac{DCS}=89.91$\pm1.77$ \newline \ac{ASD}=0.13 \newline \ac{AVD}=0.14 & \ac{DCS}=+0.24 \newline \ac{ASD}=+0.05 \newline \ac{AVD}=-0.04  \\   [0.2cm]

\cite{lou2023min} & \acs{MMS}\tablefootnote{\url{https://github.com/AngeLouCN/Min_Max_Similarity}}  & Image & Created & The classifiers extract features to create negative pairs, while projectors transform unlabeled predictions into high-dimensional features to measure pixel-wise consistency. & TransUNet & \ac{DCS}=0.920 \newline \ac{IoU}=0.861 \newline \ac{MAE}=0.021 \newline \ac{F1}=0.925 & \ac{DCS}=+0.032 \newline \ac{IoU}=+0.021 \newline \ac{MAE}=-0.01 \newline \ac{F1}=+0.012 \\  [0.2cm]

\cite{wang2021inner} & \acs{MARGAN}  & Image & Created & \acs{MARGAN} is a 3D \ac{GAN}-based approach that uses simulated training data to reduce metal artifacts in \ac{CT} images. &  marLI  & \ac{RMSE} =0.12 \newline
PSNR= 18.31 \newline
SSIM= 0.64 & \ac{RMSE}=-0.03 \newline
PSNR=+1.78 \newline
SSIM=+0.08 \\ [0.5cm]

\cite{wang2018conditional} & \acs{cGAN}  & Image & Created  & Active shape model-based method used for segmenting intra-cochlear anatomical structures. & --- & \ac{ASE}=0.173 & \ac{ASE}=-50\% \\  [0.2cm]

\cite{feng2022preservation} & \acs{MIMO}-TasNet using \acs{RNN}  & Speech & WSJ0-2mix, \newline ITA,  DEMAND  & A speech separation framework using TasNet and \ac{RNN}-\ac{EVD} to preserve spatial cues for CI users. & \acs{MIMO}-TasNet & $\Delta$\ac{ILD}=0.38 \ac{dB} & $\Delta$\ac{ILD}=-0.69 \ac{dB}  \\ 

\cite{tahmasebi2020design}&\ac{DNN} & Speech & iKala, MUSDB & Enhances music perception for \ac{CI} users, addressing pitch and timbre limitations. Objective measurements and listener experiments reveal preferences for enhanced vocals.  & --- & \ac{SDR}=8dB (with iKala) \newline \ac{SDR}=5.5dB (with MUSDB) & --- \\

\cite{lu2020speech} & Hybrid DNN  & Speech & Created  & Hearing training system consists of a speech training database, automatic lip-reading using a hybrid neural network, comparison of lip shapes, and providing a standard lip-reading sequence. & --- & \ac{Acc}=98 (gesture recognition)
\ac{Acc}=87 (lip-reading recognition) & ---  \\ [0.2cm]  

\cite{jeyalakshmi2023predicting} & \acs{ESCSO}-based LSTM  & \acs{ERP} of \acs{EEG} & Created  & Method that combines \ac{LSTM} network with \ac{ESCSO} to improve the performance of predicting \ac{CI} scores. & \acs{LSTM} & \ac{Acc}=96.03 & \ac{Acc}=+7.14  \\[0.5cm]

\cite{meeuws2020cochlear} & \ac{FOX} & Speech & Created & The study investigates the potential of utilization of \ac{AI}
in remote \ac{CI} fitting,  demonstrates the feasibility of \ac{AI}-based fitting with remote supervision. & Manual fitting & 66\% performed better with \ac{FOX} & --- \\[0.4cm]

\cite{wathour2023prospective} & \ac{FOX}  & Speech & Created & A study comparing manual and computer-assisted \ac{CI} fitting. & Manual fitting &  82\% performed better with \ac{FOX} & ---  \\ [0.4cm]

\cite{wathour2023effect} & \ac{FOX} & Speech & Created & Examined the impact of an \ac{AI}-based \ac{CI} programming tool on experienced \ac{CI} patients. The \ac{AI}-generated \ac{CI} map led to improved auditory outcomes. & Manual fitting & 89\% preferring \ac{FOX}'s map & ---  \\ 

\cite{senda2024auditory} &  {DNN and \newline  self-attention} & {Stimulus} & {Created} & {Using self-attention modules, speech reconstruction from neural activity improved by leveraging long-span neural data, outperforming \ac{CNN} and \ac{MLP} models, enhancing hearing evaluation and neurosurgery.} &  {\ac{CNN} and \ac{MLP}} & {ESTOI=0.239}    &  {ESTOI=+0.009}  \\ [0.5cm]

\cite{kassjanski2024automated} &  {Bi-LSTM} & {Audiograms} & Created & A \ac{Bi-LSTM} {classifies tonal audiometry data, aiding general practitioners and reducing audiologists' burden by diagnosing hearing loss types independently.} &  {\ac{LSTM}}  & {\ac{Acc}=99.33\% }   &  {\ac{Acc}=+1.04\%}  \\ [0.5cm]

\cite{jehn2024cnns} &  {CNN and SVM} & Speech & Created & {A \ac{CNN} significantly improves \ac{AAD} for CI users, enhancing listening experiences by steering auditory prostheses towards target speech.} & --- & {\ac{Acc}=74\%} & --- \\ [0.3cm]

%\cite{borjigin2024deep} &  {RNN}  &{Speech} & {LibriSpeech, WHAM} & { \ac{RNN} and SepFormer, significantly enhance speech intelligibility for \ac{CI} users in noisy environments, outperforming conventional noise reduction strategies.} & {DCCRN, DCCTN} & --- & --- \\ [0.2cm]
 \hline
\end{tabular}
}
\begin{flushleft}
    Abbreviations: Project link (PL), Not available (NA), Compared to other \ac{DL} methods (CTODLM), \Ac{ESTOI}.
\end{flushleft}
\end{table*}

\subsection{\acs{GAN}-based methods}

\begin{figure*}[ht!]
    \centering
    \includegraphics[scale=0.6]{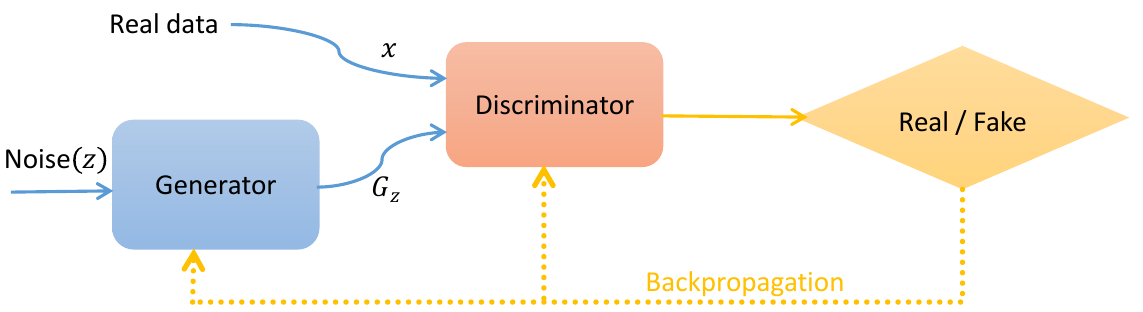}
    \caption{ Illustrative depiction of a standard \ac{GAN}.}
    \label{fig:iGAN}
\end{figure*}

A \ac{GAN} is a type of \ac{AI} model consisting of two neural networks, a generator, and a discriminator, engaged in a competitive learning process as presented in Figure \ref{fig:iGAN}.

The generator aims to create realistic data, such as images, while the discriminator tries to differentiate between real and generated samples. This adversarial training dynamic leads to the refinement of the generator's output, generating increasingly authentic data. The objective is for the generator to produce data that is indistinguishable from real samples. The training process is represented by the minimax game framework, with the \ac{GAN} objective function given by:

\begin{equation}
\begin{split}
  \min_G \max_D V(D, G) = & \, \mathbb{E}_{x \sim p_{\text{data}}(x)} [\log D(x)] \\
  & + \mathbb{E}_{z \sim p_z(z)} [\log (1 - D(G(z)))]
\end{split}
\end{equation}

In the \ac{GAN} objective function, \(\mathbb{E}_{x \sim p_{\text{data}}(x)}\) and \(\mathbb{E}_{z \sim p_z(z)}\) indicate the expected values over real data samples \(x\)  and noise samples \(z\), respectively. \(G\) generates samples, \(D\) discriminates between real and generated samples, \(p_{\text{data}}\) and \(p_z\) are the distributions of real data and noise, respectively. Using \ac{GAN}, the research in \cite{wang2019deep} proposes a \ac{DL}-based method for reducing metal artifacts in post-operative \ac{CT} imaging. The method utilizes a 3D-\ac{GAN} trained on a large number of pre-operative images with simulated metal artifacts. The \ac{GAN} generates artifact-free images by reducing the metal artifacts. The effectiveness of the method is evaluated quantitatively and qualitatively, showing promising results compared to classical artifact reduction algorithms. The approach overcomes the challenges of post-operative assessment of cochlear implantation caused by metal artifacts, and it does not require registration of pre and post-operative images. The 3D-\ac{GAN} improves spatial consistency and is applicable to various types of artifacts. In addition, Wang et al. in theirs  paper \cite{wang2021inner},  proposes a 3D metal artifact reduction algorithm for post-operative high-resolution \ac{CT} imaging. The algorithm is based on a \ac{GAN} that uses simulated physically realistic \ac{CT} metal artifacts created by \ac{CI} electrodes. The generated images are used to train the network for artifact reduction. The  metal artifact reduction-\ac{GAN} based  method as described in \cite{wang2021inner}, utilizes a three-step process for reducing metal artifacts. Firstly, a simulation is performed to replicate \ac{CI} positioning. Secondly, a physical simulation of CI metal artifacts is conducted. Lastly, a 3D GAN is trained using both simulated and preoperative datasets. The generator component of the GAN generates an image that has reduced metal artifacts, while the discriminator network is responsible for determining whether the input image contains metal artifacts or not. The method was evaluated on clinical \ac{CT} images of \ac{CI} postoperative cases and outperformed other general metal artifact reduction approaches. The paper introduces a novel approach that combines the physical simulation of metal artifacts with 3D-\ac{GAN}, providing a promising solution for improving the visual assessment of post-operative imaging in \ac{CT}. 

Similarly, for \ac{CI} metal artifacts reduction also, a \ac{cGAN} were proposed by Wang et al. \cite{wang2018conditional}. The approach involves training a \ac{cGAN} to learn mapping from artifact-affected \acp{CT} to artifact-free \acp{CT}. During inference, the \ac{cGAN} generated \ac{CT} images with removed artifacts. Additionally, a band-wise normalization method was proposed as a preprocessing step to improve the performance of \ac{cGAN}. The method was evaluated on post-implantation \acp{CT} recipients, and the quality of the artifact-corrected images was quantitatively assessed using \ac{P2PE}. The results demonstrate promising artifact reduction, outperforming the previously proposed techniques. The authors evaluates the quality of artifact-corrected images using a quantitative metric based on segmentations of intracochlear anatomical structures. Specifically, the segmentation results obtained from a previously published method were compared between real preimplantation \acp{CT} and artifact-corrected \acp{CT} generated by the proposed method. The \ac{ASE} was used as a metric to assess the accuracy of the segmentation. The paper reports that the proposed method achieves an \ac{ASE} of 0.18 mm, which is approximately half of the error obtained with a previously proposed technique. {Gogate et al. \cite{gogate2024robust} propose a robust real-time audio-visual speech enhancement framework for \acp{CI}. By leveraging a \ac{GAN} and \ac{DNN}, the framework effectively addresses visual and acoustic speech noise in real-world environments. Experimental results demonstrate significant improvements in speech quality and intelligibility, offering potential benefits for \ac{CI} users in noisy social settings.}

\subsection{\acs{RNN}/\acs{LSTM}-based methods}

\Acp{RNN} are a class of artificial neural networks designed for sequential data processing. They maintain hidden state information that is updated at each time step, allowing them to capture temporal dependencies. The hidden state at time \( t \), denoted as \( h_t \), is computed based on the input \( x_t \), the previous hidden state \( h_{t-1} \), and model parameters \( W \) and \( U \), with \( b \) representing the bias term. The equations governing the hidden state update are given by:

\begin{equation}
 h_t = \sigma(Wx_t + Uh_{t-1} + b)   
\end{equation}

\noindent where, \( \sigma \) is an activation function, typically the hyperbolic tangent or \ac{ReLU}. Several speech processing techniques that utilize \ac{CI} are based on \acp{RNN}.

\ac{CI} users struggle with music perception, and many studies have shown that enhancing music vocals improves their enjoyment. The study described by Gajęcki. et al.  \cite{gajkecki2018deep} explores source separation algorithms to remix pop songs by emphasizing the lead-singing voice. \Ac{DCAE}, \ac{DRNN}, \ac{MLP}, and \ac{NMF} were evaluated through perceptual experiments involving \ac{CI} recipients and normal hearing subjects. The results show that a \ac{MLP} and \ac{DRNN} perform well, providing minimal distortions and artifacts that are not perceived by \ac{CI} users. The paper also highlights the benefits of  the implementation of a \ac{MLP} for real-time audio source separation to enhance music for \ac{CI} users due to their reduced computation time.
In addition, The study described in \cite{feng2022preservation}  proposes a speech separation framework for \ac{CI} users using TasNet and \ac{RNN}-\ac{EVD}. TasNet, a non-causal \ac{MIMO}-based method, is employed as the speech separation module. \ac{RNN}-\ac{EVD}, which combines \acp{RNN} with \ac{EVD}, is utilized to preserve spatial cues. The framework aims to effectively separate speech and reduce \ac{ILD} errors. The \ac{RNN}-\ac{EVD} network is trained using $\Delta$\ac{ILD} as the objective, and an additional \ac{SNR} term is added to the loss function for convergence. The experimental results demonstrate the effectiveness of the proposed framework in preserving \ac{ILD} cues for \ac{CI} users in various hearing scenarios. {Borjigin et al. \cite{borjigin2024deep} explores the use of \ac{DNN} algorithms, specifically \ac{RNN} and SepFormer, a Transformer-based algorithm, in speech separation applications to improve speech intelligibility for \ac{CI} users in multi-talker interference. The algorithms were trained with a customized dataset and tested with thirteen \ac{CI} listeners. Both \ac{RNN} and SepFormer significantly improved speech intelligibility in noise without compromising speech quality, indicating the potential of \ac{DNN} algorithms as a solution for multi-talker noise interference.}

The \acf{LSTM}, an enhanced version of the \ac{RNN}, addresses limitations observed in \acp{RNN} under specific conditions  \cite{kheddar2023deep,hochreiter1997long}. Unlike \ac{RNN}, \ac{LSTM} excels in preserving past information, making it suitable for tasks with long-term dependencies. Comprising \ac{LSTM} units forming layers, each unit regulates information flow through input, output, and forget gates, allowing for prolonged retention of crucial information. The forward pass equations (\ref{lstm}) illustrate this process \cite{kheddar2023deep}. The symbols $L_i$ and $L_j$ denote input and output, while $A_f$, $A_i$, and $A_j$ represent activation vectors for forget, input, and output gates. $V_c$ is the cell state vector, and $\sigma$ for the sigmoid activation function and $\odot$ for element-wise multiplication. This \ac{LSTM} structure with weight matrices $W$ and $U$ and bias vector $b$ is outlined by \cite{greff2016lstm}.

\begin{align}
\begin{split}
A_f &= \sigma(W_f * L_i + U_f * L_{j-1} + b_f), \\  A_i &  = \sigma(W_i * L_i + U_i * L_{j-1} + b_i), \\
L_j &= A_j \odot \tanh(V_c), \\
A_j &= \sigma(W_0 * L_i + U_0 * L_{j-1} + b_j), \\
V_c &= A_f * V_{c-1} + A_i * \text{cell\_state}(W_c * L_i + U_c * L_{j-1} + b_c).
\label{lstm}
\end{split}
\end{align}

\begin{figure}[ht!]
    \centering
    \includegraphics[scale=0.7]{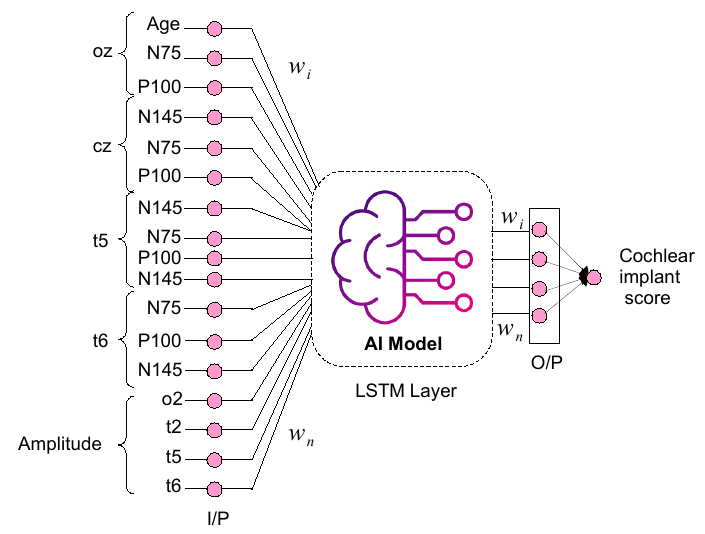}
    \caption{ The design of the LSTM architecture suggested by  Jeyalakshmi et al. \cite{jeyalakshmi2023predicting}. }
    \label{fig:LSTM1}
\end{figure}

Recently, several schemes for \ac{CI} and utilizing LSTM have been proposed in the literature. The study described by Lu. et al. in 2020 \cite{lu2020speech} introduces a speech training system designed for individuals with hearing impairments, such as those with \acp{CI}, as well as individuals with dysphonia, utilizing automated lip-reading recognition. The system combines \ac{CNN} and \ac{RNN} to compare mouth shapes and train speech skills. It includes a speech training database, automatic lip-reading using a hybrid neural network, matching lip shapes with sign language vocabulary, and drawing comparison data. The system enables hearing-impaired individuals to analyze and improve their vocal lip shapes independently. It also supports the use of medical devices for correct pronunciation. Experimental results demonstrate the system's effectiveness in correcting lip shape and enhancing speech ability. The proposed model utilizes ResNet50, MobileNet, and \ac{LSTM} networks for accurate lip-reading recognition. Later on, the scientific paper published by Chu et al. in 2021 \cite{chu2021causal} proposes a causal \ac{DL} framework for classifying phonemes in \acp{CI} to enhance speech intelligibility. The authors trained \ac{LSTM} networks using features extracted at the time-frequency resolution of a \ac{CI} processor. They compared \ac{CI}-inspired features (log STFT power spectrum, log \ac{ACE} power spectrum, and log-mel-filterbank) with traditional \ac{ASR} features. The results showed that \ac{CI}-inspired features outperformed traditional features, providing slightly higher levels of performance. The author claimed that, this study is the first to introduce a classification framework with the potential to categorize phonetic units in real-time in a \ac{CI}, offering possibilities for improving speech recognition in reverberant environments for \ac{CI} users. In 2023, Huang et al. proposed in \cite{huang2023electrodenet}, a \ac{DL}-based sound coding strategy for \acp{CI}, called ElectrodeNet. By leveraging \ac{DNN}, \ac{CNN}, and \ac{LSTM}, ElectrodeNet replaces conventional envelope detection in the \ac{ACE} strategy. Objective evaluations using measures like \ac{STOI} and \ac{NCM} demonstrate strong correlations between ElectrodeNet and \ac{ACE}. Additionally, subjective tests with normal-hearing listeners confirm the effectiveness of ElectrodeNet in sentence recognition for vocoded Mandarin speech. The study extends ElectrodeNet with ElectrodeNet-CS, incorporating \ac{CS} through a modified \ac{DNN} network. ElectrodeNet-CS produces N-of-M compatible electrode patterns and performs comparably or slightly better than \ac{ACE} in terms of \ac{STOI} and sentence recognition. This research showcases the feasibility and potential of deep learning in \ac{CI} coding strategies, paving the way for future advancements in \ac{AI}-powered \ac{CI} systems. Similarly, the research presented by Jeyalakshmi et al. \cite{jeyalakshmi2023predicting} focuses on predicting \ac{CI} scores for children aged 5 to 10 using a reconfigured \ac{LSTM} network as illustrated in Figure \ref{fig:LSTM1}. The proposed architecture aims to enhance language development skills in children with auditory deprivation, this could be achieved by guiding \ac{CI} programming through the analysis of cross-modal data obtained from previously programmed patients. The research utilizes visual cross-modal plasticity and visual evoked potential to discover patterns in the data that can predict outcomes for future patients. The proposed methodology involves the use of \ac{LSTM} network and \ac{ESCSO} to identify optimal weights. The results demonstrate the superiority of the \ac{ESCSO}-based \ac{LSTM} technique over other methods.

In Figure \ref{fig:LSTM1}, "oz," "cz," "t5," and "t6" refer to specific electrode placements or positions on the scalp in the international 10-20 system for \ac{EEG} or \ac{ERP} recordings. These positions represent specific areas on the scalp where electrodes are attached to measure electrical activity in the brain. The amplitude represents the intensity or strength of the electrical signal detected at a particular point on the scalp, reflecting the neural activity in the corresponding brain region. The parameters "N75," "P100," and "N145" refer to specific components or peaks of \acp{ERP} obtained from \ac{EEG} recordings. \acp{ERP} are electrical responses recorded from the brain in response to specific stimuli or events, and they reflect the neural processing associated with those stimuli. Besides, \( I/P \) represent inputs, \( O/P \) for output, and \(W\) represent Weights. {Recently, \cite{kassjanski2024automated} proposes a neural network model based on \ac{Bi-LSTM} architecture for classifying hearing loss types using tonal audiometry data. The model achieves 99.33\% classification accuracy on external datasets. The system can assist general practitioners in independently classifying audiometry results, reducing the burden on audiologists and improving diagnostic accuracy. The study, that may help assisting \ac{CI} patients,  aims to surpass the current \ac{SOTA} accuracy rate of 95.5\% achieved through \ac{DT}.}

\subsection{\acs{AE}-based methods}
An \ac{AE} is a type of neural network designed for unsupervised learning, tasked with encoding input data into a compressed representation and decoding it back to the original form. Examples include \acp{VAE}, which balance data compression with generative modeling, \acp{CAE}, which employ convolutional layers for efficient feature learning and reconstruction, and sparse \acp{AE}, which induce sparsity, promoting selectivity in feature representation, among others. The encoding equation typically involves a mapping function, such as \(h = f(x)\), where \(h\) is the encoded representation and \(x\) is the input. The decoding equation is the reconstruction of the input, often expressed as \(r = g(h)\), where \(r\) is the reconstructed output and \(g\) is the decoding function. \acp{AE} find applications in data compression, denoising, feature learning, and more. Recently, many research papers for \ac{CI} have been proposed that are based on \ac{AE}. 

As a point of the case, the scientific paper \cite{lai2016deep} delves into the pivotal objective of enhancing speech perception for \ac{CI} users in noisy conditions, recognizing the critical role of \ac{NR} in this pursuit. The proposed method, named \ac{DDAE}-\ac{NR}, has been proven effective in restoring clean speech. The study focuses on evaluating the \ac{DDAE}-based \ac{NR} using envelope-based vocoded speech, mimicking \ac{CI} devices.
The procedure of \ac{DDAE}-based \ac{NR} can be split into two main stages: training and testing. During the training phase, a collection of pairs of noisy and clean speech signals is prepared. These signals are initially transformed into the frequency domain using an \ac{FFT}. The logarithmic amplitudes of the noisy and clean speech spectra are then used as inputs and outputs, respectively, for the \ac{DDAE} model.

Key findings underscore the superior intelligibility of \ac{DDAE}-based \ac{NR} in vocoded speech compared to \ac{SOTA} conventional methods, indicating its potential implementation in \ac{CI} speech processors. However, the study acknowledges the use of noise-vocoded speech simulation for evaluation and emphasizes the need for further validation with real \ac{CI} recipients in clinical settings, addressing potential inconsistencies in the transition to actual \ac{CI} devices.

A zero-delay  \ac{DAE} is proposed in \cite{hinrichs2022vector} for compressing and transmitting electrical stimulation patterns generated by \acp{CI}. The goal is to conserve battery power in wireless transmission while maintaining low latency, which is crucial for speech perception in \ac{CI} users. The \ac{DAE} architecture is optimized using Bayesian optimization and the \ac{STOI}. The results show that the proposed \ac{DAE} achieves equal or superior speech understanding compared to audio codecs, with reference vocoder \ac{STOI} scores at 13.5 kbit/s. This approach offers a promising solution for efficient and real-time compression of \ac{CI} stimulation patterns, addressing the constraints of low latency and battery power consumption. Moreover, The research in \cite{chen2023deeply} focuses on achieving accurate segmentation of the vestibule in \ac{CT} images, a crucial step for clinical diagnosis of congenital ear malformations and \acp{CI}. The challenges addressed include the small size and irregular shape of the vestibule, making segmentation difficult, and the limited availability of labelled samples due to high labour costs. To overcome these challenges, the proposed method introduces a vestibule segmentation network within a basic encoder-decoder framework. Key innovations include the incorporation of a \ac{Res-CA} block for channel attention, a \ac{GCPFE} module for global context information, an , \ac{ACE-Loss} function for detailed boundary learning, and a \ac{DS} mechanism to enhance network robustness. The network architecture utilizes ResNet34 as the backbone with skip connections for multi-level feature fusion.  Results showcases a high performance, and are supported by comprehensive comparisons, ablation studies, and visualized segmentation outcomes. The study also acknowledges limitations, such as reliance on professional annotations.

In addition, the study presented in \cite{fan2021hybrid} aims to enhance the accuracy and robustness of \ac{ICA} segmentation, a vital component in preoperative decisions, insertion planning, and postoperative adjustments for \ac{CI} procedures. The \ac{ICA} includes structures such as \ac{ST}, \ac{SV}, and the \ac{AR}. The researchers employed two segmentation methods, \ac{ASM} and \ac{DL} based on 3D U-Net \ac{AE}, and combined them to achieve improved accuracy and robustness. A two-level training strategy involved pretraining on clinical \acp{CT} using \ac{ASM} and fine-tuning on specimens' \acp{CT} with ground truth.  Results demonstrated that \ac{DL} methods outperformed \ac{ASM} in accuracy. While a trade-off between accuracy and robustness was observed, the combined \ac{DL} and \ac{ASM} approach showed improvements in both aspects. The study concludes that the proposed \ac{DL} and \ac{ASM} method effectively balances accuracy and robustness for \ac{ICA} segmentation, highlighting the potential of \ac{DL}-based methods, especially when integrated with \ac{ASM}, to enhance \ac{CI} procedures.

The proposed \ac{MMS} methodology in \cite{lou2023min} represents a groundbreaking approach to semi-supervised segmentation networks, particularly in the context of medical applications such as endoscopy surgical tool segmentation and \ac{CI} surgery. \ac{MMS} is introduced through dual-view training with contrastive learning, utilizing classifiers and projectors to create negative, positive, and negative pairs. The inclusion of pixel-wise contrastive loss ensures the consistency of unlabeled predictions. In the evaluation phase, \ac{MMS} was tested on four public endoscopy surgical tool segmentation datasets and a manually annotated \ac{CI} surgery dataset. The results demonstrate its superiority over \ac{SOTA} semi-supervised and fully supervised segmentation algorithms, both quantitatively and qualitatively. Notably, \ac{MMS} exhibited successful recognition of unknown surgical tools, providing reliable predictions, and achieved real-time video segmentation with an impressive inference speed of about 40 frames per second. This signifies the potential of \ac{MMS} as a highly effective and efficient tool in medical image segmentation, showcasing its applicability in real-world surgical scenarios. 

\begin{figure*}[h!]
    \centering
    \includegraphics[scale=1]{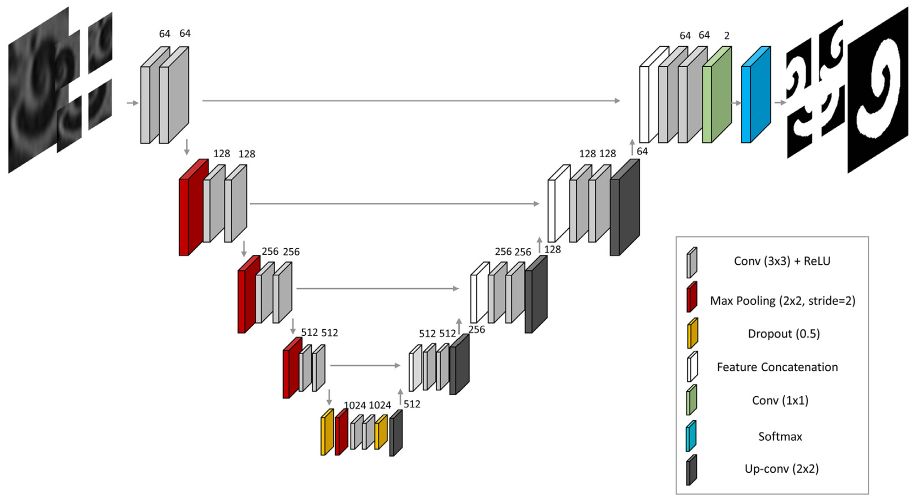}
    \caption{ Encoder-decoder network used in the pixel-wise classification model \cite{heutink2020multi}.}
    \label{fig:AE2_2}
\end{figure*}

Similarly, The primary aim of the study in \cite{heutink2020multi} is to devise an automated method for the segmentation and measurement of the human cochlea in \ac{UHR} \ac{CT}-images. The objective is to explore variations in cochlear size to enhance outcomes in cochlear surgery through personalized implant planning. Initially, the input scans undergo a two-step process using a detection module and a pixel-wise classification module for cochlea localization and segmentation, respectively using an \ac{AE} as illustrated in Figure \ref{fig:AE2_2}. The detection module reduces the search area for the classification module, improving algorithm speed and reducing false positives. Both modules are trained on image patches, allowing for a larger training set size by generating multiple examples from each scan. The segmented cochlear structure then proceeds to a final module that combines \ac{DL} and thinning algorithms to extract patient-specific anatomical measurements. \ac{DL} is employed in each step to leverage its ability to learn directly from input data, providing automatic results without the need for user-adjustable parameters during testing.

\subsection{RL-based methods }

\begin{figure}[h]
    \centering
    \includegraphics[scale=0.85]{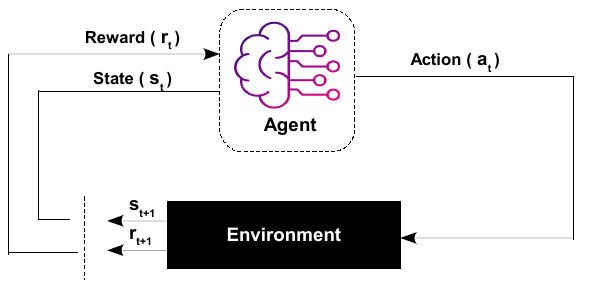}
    \caption{RL principle.}
    \label{fig:rl}
\end{figure}

\Ac{DRL} is a subfield of \ac{ML} that enable agents to learn and make decisions in complex environments. It involves training an agent to interact with an environment, learn from the outcomes of its actions, and optimize its behavior over time \cite{guerianiRL2024}. In traditional \ac{RL}, agents learn by trial and error, receiving feedback in the form of rewards or penalties for their actions. However, \ac{DRL} incorporates \ac{DNN}, make it capable of learning complex patterns and representations from raw data. This allows \ac{DRL} agents to handle high-dimensional input spaces, such as images or sensor data, and make more sophisticated decisions. Figure \ref{fig:rl} illustrate the priciple of RL.

The paper presented by radutoiu et al. \cite{radutoiu2022accurate} presents a novel method for accurately localizing \acp{ROI} in the inner ear using \ac{DRL}. The proposed method addresses the challenges of robust \ac{ROI} extraction in full head \ac{CT} scans, which is crucial for \ac{CI} surgery. The approach utilizes communicative multi-agent \ac{RL} and landmarks specifically designed to extract orientation parameters. The method achieves an average estimated error of 1.07 mm for landmark localization. The extracted \acp{ROI} demonstrate an \ac{IoU} of 0.84 and a dice similarity coefficient of 0.91, conducted over 140 full head CT scans, showing promising results for automatic \ac{ROI} extraction in medical imaging. 
In addition, Lopez et al. presents in \cite{lopez2021facial} a pipeline for characterizing facial and cochlear nerves in \ac{CT} scans using \ac{DRL}. Key landmarks around these nerves are located using a communicative multi-agent \ac{DRL} model. The pipeline includes automated measurement of the cochlear nerve canal diameter, extraction and segmentation of the cochlear nerve cross-section, and path selection for the facial nerve characterization. The pipeline was developed and evaluated using 119 clinical \ac{CT} images. The results show accurate characterizations of the nerves in the cochlear region, providing reliable measurements for computer-aided diagnosis and surgery planning. The proposed approach demonstrates the potential of \ac{DRL} for landmark detection in challenging medical imaging tasks.

\section{Applications of DL-based medical \ac{CI}}
\label{sec4}

\begin{figure*}
    \centering
    \includegraphics[scale=0.85]{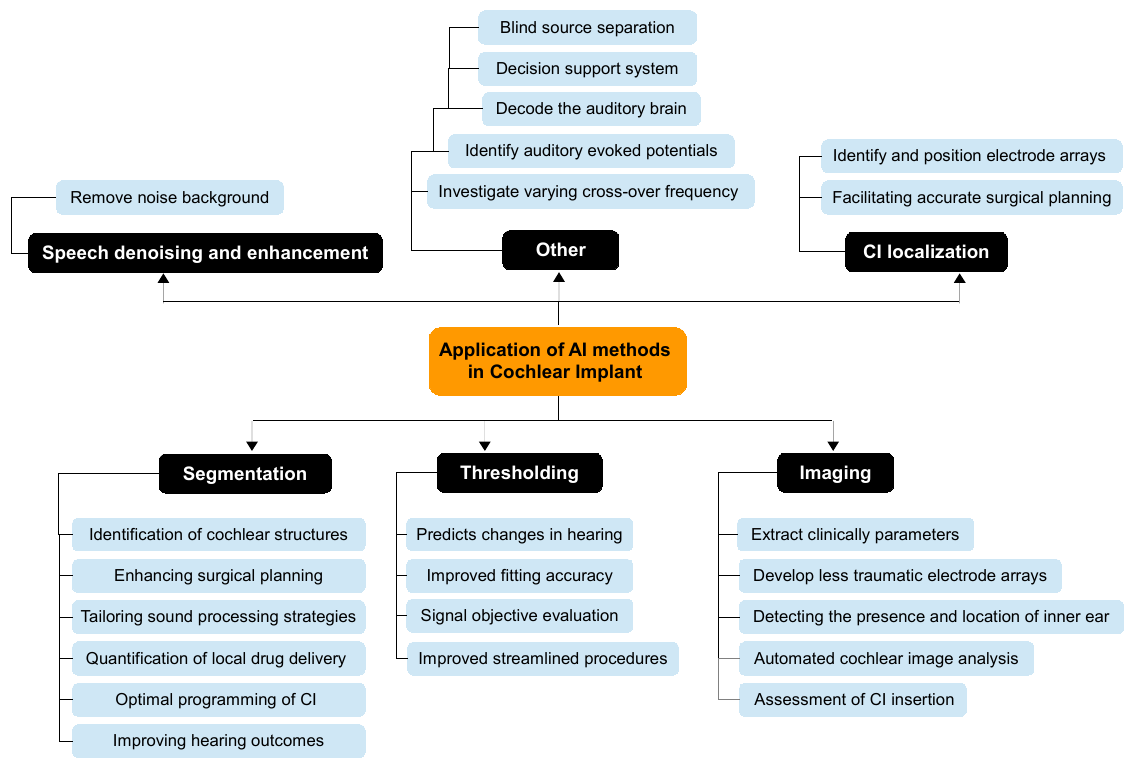}
    \caption{Taxonomy of AI-based applications for \acp{CI} and their benefits.}
    \label{fig:tax-apps}
\end{figure*}

This section explores the application of deep learning in the field of cochlear implants, encompassing tasks such as speech denoising and enhancement, segmentation for precise identification and analysis of cochlear structures, thresholding, imaging, localization of \ac{CI}, and more. Figure \ref{fig:tax-apps} provides a comprehensive overview of AI-based applications for \acp{CI} and theirs associated benefits. Furthermore, Table \ref{tab:Apps} summarizes various applications based on AI techniques, highlighting their performance, and  pros and cons.

\subsection{Speech denoising and enhancement}

The integration of  \ac{ML}, and \ac{DL} has proven invaluable in the field of \acp{CI}. Researchers have harnessed these technologies to tackle numerous challenges and enhance speech perception for individuals with hearing impairments. The work \cite{wang2017deep,lai2018deep} employed \ac{DDAE} approach to reduce unwanted background noise in speech signals. However,  Lai et al. in \cite{lai2018deep} devised a \ac{NR} system that employed a noise classifier and \ac{DDAE}, specifically tailored for Mandarin-speaking \ac{CI} recipients. The proposed schemes \cite{gajecki2023deep, gajecki2022end, healy2015algorithm} aim to perform end-to-end speech denoising, with the goal of enhancing speech intelligibility in noisy environments. Gajecki et al. \cite{gajecki2023deep, gajecki2022end} employed \ac{DNN} to develop the Deep ACE method, while Healy et al. in \cite{healy2015algorithm} utilized \ac{DNN} to separate speech from background noise.  These examples underscore the broad spectrum of applications of \ac{AI}, particularly \ac{DL} techniques, in addressing challenges related to noise reduction and enhancing speech intelligibility in \acp{CI} applications.

Moreover, Kang et al. \cite{kang2021deep} used \ac{DL}-based speech enhancement algorithms to optimize speech perception for \ac{CI} recipients. Their approach achieved a balance between noise suppression and speech distortion by experimenting with different loss functions. Hu et al. \cite{hu2010environment} developed environment-specific noise suppression algorithms for \acp{CI} using \ac{ML} techniques. They improved the processed sound by classifying and selecting envelope amplitudes based on the \ac{SNR} in each channel. Banerjee et al. \cite{banerjee2015identifying} employed online unsupervised algorithms to learn features from the speech of individuals with severe-to-profound hearing loss, aiming to enhance the audibility of speech through modified signal processing. Li et al. \cite{li2021improved} developed an improved \ac{NR} system for \acp{CI} using \ac{DL}, specifically \ac{DDAE}, and knowledge transfer technology. Their goal was to enhance speech intelligibility in noisy conditions. Fischer et al. \cite{fischer2021speech} utilized \ac{DL}-based virtual sensing of head-mounted microphones to improve speech signals in cocktail party scenarios for individuals with hearing loss, resulting in enhanced speech quality and intelligibility, particularly in noisy environments. These studies exemplify the versatility of \ac{AI} and \ac{DL} in addressing various challenges associated with \acp{CI}, including \ac{NR}, speech enhancement, and improved speech perception. Furthermore, the paper by Chu et al. \cite{chu2018using} explores the application of \ac{ML} algorithms to mitigate the effects of reverberation and noise in \acp{CI}, to improve speech intelligibility for individuals with severe hearing loss.

\begin{table*}[h!]
\caption{Summary of the performance and limitations of specific DL applications dedicated to \acp{CI}. In cases where multiple tests are conducted, only the best performance is reported. } 
\label{tab:Apps}
\scriptsize
\begin{tabular}{m{0.5cm}m{0.5cm}m{1cm}m{5cm}m{2cm}m{5.5cm}m{0.5cm}}
\hline
Ref. & Year & DLM & Description & BP  & Limitations & PLA  \\ \hline
\midrule

\cite{li2021improved} &  2021 &  NC and DDAE\_T  & The \ac{DL} model comprises Siren noise at 6dB, a classifier, and the \ac{DDAE}. The \ac{TL} is incorporated to help reduce the number of parameters in the model. &  PESQ= 3.037 \newline STOI= 0.843 &  The performance of the \ac{DL} model can be degraded when there is a mismatch between the training and testing data. & No\\[0.5cm]

\cite{gajecki2023deep} &  2023 &  DeepACE  &  Takes the raw audio signal as input and generates a denoised electrodogram, which represents the electrical stimulation patterns applied to the electrodes over time.   &  STOI =0.807  &  
Heavy data reliance impedes performance with insufficient or biased data, posing generalization challenges. Computational complexity, interpretability issues, and vulnerability are concerns.
 & Yes\tablefootnote{\url{https://github.com/APGDHZ/DeepACE2.0}} \\[0.5cm]

\cite{fischer2021speech} &  2021  &  U-Net  &  The proposed approach involves using deep virtual sensing to estimate microphone signals, improving speech quality in cocktail party scenarios for hearing aid and \ac{CI} users.  &  \acs{PESQ}=2, \ac{STOI}=0.53, SI-\ac{SDR}=-28.91dB  &  
 Challenges include long delay impracticality for hearing aids, binaural cues omission, insufficient network optimization, and exploring robustness in high-reverberation environments.
& No \\[0.5cm]

\cite{margeta2022web} &  2022 & \ac{CNN}  &  A web-based research platform called Nautilus, which utilizes \ac{DL} techniques for automated image processing in cochlear implantation-related studies.  &  \ac{DCS}= 80 $\pm3$\%, \ac{ASD}= 0.19 $\pm0.04$ mm  & 
Dependence on large amounts of annotated training data poses a challenge. Ensuring generalizability across various imaging conditions and populations is crucial.
 & Yes\tablefootnote{\href{nautilus_info@oticonmedical.com}{nautilus\_info@oticonmedical.com} \textit{available upon reasonable request}}\\[0.5cm]

\cite{chen2023deeply} &  2023  &  \ac{AE}+\ac{Res-CA}  &  Is a vestibule segmentation network for CT images. It is based on the basic encoder-decoder framework and incorporates \ac{Res-CA} Block, \ac{GCPFE} and \ac{ACE-Loss} Function &  \ac{DCS}=94.77 $\pm 2.45$\%, \ac{ASD}=0.06 mm
  & 
Limited training data, small object handling, generalization capability, and computational resource constraints are areas of concern.
 & No  \\[0.5cm]

 \cite{chi2019deep} &  2019  &  \ac{cGAN}  &  Is a \ac{DL}-based method for accurately localizing \ac{CI} electrode contacts in \ac{CT} images, facilitating customization of \ac{CI} settings.  &  success rate of localization =96.7 
  &  coarse resolution, metal artifacts, difficulty isolating contacts from cortical bones, site-dependent image quality, and variability among electrode array manufacturers.  & No  \\[0.5cm]

\cite{radutoiu2022accurate} & 2022 & \ac{DRL} & Is a novel method for accurately extracting \acp{ROI} in the inner ear from \ac{CT} scans. The approach achieves high precision and demonstrates promising results for surgical planning in cochlear implantation. & \ac{IoU}= 0.84 \newline DSC=0.91 & Small dataset, image quality variability, and computational requirements not specified. & No \\[0.5cm]

\cite{heutink2020multi} & 2020 & U-Net & Is a multi-scale \ac{DL} framework for automatic segmentation and measurement of the human cochlea in clinical ultra-high-resolution \ac{CT} images, with potential applications in personalized \ac{CI} surgery. & DSC=0.90 $\pm0.03$ \newline BF=0.95$\pm0.03$ & The work include a small dataset, potential variability in image quality, lack of external validation, and limited assessment of clinical utility and computational requirements. & No \\[0.5cm]

\cite{li2023application} & 2023 & UNETR & Explores the feasibility of using a \ac{DL} method based on the UNETR model for automatic segmentation of the cochlea in temporal bone \ac{CT} images. & DSC=0.92 & The small dataset used, the variability in image quality, and the absence of specifications regarding computational requirements. & No \\[0.5cm]

\cite{zhang2019two} & 2019 & 3D U-NET & Is a two-level training approach using a \ac{DL} method to accurately segment the intra-cochlear anatomy in head \ac{CT} scans. The method combines an active shape model-based method and a 3D U-Net model & DSC=0.87 &Limited dataset, variable image quality, lack of external validation, limited assessment of clinical utility, and no specifications on computational requirements. & No \\[0.5cm]

\cite{schuerch2023objective} & 2023 & AlexNet & The study assesses repeatability, thresholds, and tonotopic patterns using a DL-based algorithm, providing insights into inner ear function and potential clinical applications. & \ac{Acc}=83.8\% & Potential dependence on the quality of the input data, limited generalizability to different patient populations or implant systems, and the need for further external validation and comparison with expert visual inspection. & No \\[0.5cm]

\cite{jehn2024cnns} &  { 2024} &  {CNN}  & {The model explores the use of \acp{CNN} to improve the decoding of selective attention to speech in \ac{CI} users, aiming to enhance their listening experience in challenging environments.} &  {Acc= 74\%} &  {Small sample size of 25 CI users, limiting the generalizability of the findings, and the presence of electrical artifacts in \Ac{EEG} recordings caused by the implant potentially affecting the accuracy of decoding.} & Yes\tablefootnote{\url{https://github.com/Constantin-Jehn/CNN_AAD_CI_Users}}\\[0.5cm]
  
\hline
\end{tabular}
\begin{flushleft}
Abbreviations: Deep learning model (DLM); Best performance (BP); Project link availability (PLA).
\end{flushleft}
\end{table*}

\subsection{Imaging}

\ac{DL} methods have revolutionized \ac{CI} applications by leveraging imaging data for enhanced analysis and optimization. Hussain et al. \cite{hussain2023anatomical} employed image analysis tools, such as the oticon medical nautilus software, to automatically detect landmarks and extract clinically relevant parameters from cochlear \ac{CT} images. This approach provides valuable insights into cochlear morphology, facilitating the development of less traumatic electrode arrays for cochlear implantation. Zhang et al. \cite{zhang2018automatic} focused on automatically detecting the presence and location of inner ears in head \ac{CT} images, aiming to assist in image-guided \ac{CI} programming for patients with profound hearing loss. Regodic et al. \cite{regodic2020automatic} introduced an algorithm that utilizes a \ac{CNN} for automatic fiducial marker detection and localization in \ac{CT} images, enhancing registration accuracy, reducing human errors, and shortening intervention time in computer-assisted surgeries. Margeta et al. \cite{margeta2022web} presented Nautilus, a web-based research platform that employs \ac{AI} and image processing techniques for automated cochlear image analysis. This platform enables accurate delineation of cochlear structures, detection of electrode locations, and personalized pre- and post-operative metrics, facilitating clinical exploration in cochlear implantation studies. Li et al. \cite{li2020clinical} proposed the integration of \ac{DL} techniques into a clinical $\mu$\ac{CT} system to optimize imaging performance, improve reconstruction accuracy, and enhance diagnostic capabilities in temporal bone imaging and other clinical applications. Wang et al. \cite{wang2019deep} addressed the reduction of metal artifacts in post-operative \ac{CI} \ac{CT} imaging using a 3D \ac{GAN}, enabling better analysis of electrode positions and assessment of \ac{CI} insertion. These advancements highlight the significant role of \ac{DL}, \ac{ML}, and \ac{AI} in leveraging imaging data for improved \ac{CI} analysis, design, and surgical procedures.

In addition to the previous advancements, \ac{DL} and \ac{AI} have been applied to various aspects of \ac{CI} applications using imaging data.  Chen et al. \cite{chen2023deeply} utilize \ac{AI} for accurate vestibule segmentation in \ac{CT} images, which plays a crucial role in the clinical diagnosis of congenital ear malformations and \ac{CI} procedures. Kugler et al. \cite{kugler2020i3posnet} employ \ac{AI} techniques to accurately estimate instrument pose from X-ray images in temporal bone surgery, enabling high-precision navigation and facilitating minimally invasive procedures. Waldeck et al. \cite{waldeck2022new} develop an ultra-fast algorithm that utilizes automated cochlear image registration to detect misalignment in \acp{CI}, significantly reducing the time required for diagnosis compared to traditional multiplanar reconstruction analysis. Finally, Chen et al. \cite{chen2009three} focus on creating a three-dimensional finite element model of the brain based on \ac{MRI} data to analyze and optimize the current flow path induced by \acp{CI}. This application of \ac{AI} contributes to the improvement of implant design in the future. These innovative approaches demonstrate the diverse applications of \ac{DL}, \ac{ML}, and \ac{AI} in \ac{CI} research, ranging from scene understanding to precise segmentation, instrument pose estimation, misalignment detection, and implant design optimization.

\subsection{Segmentation}

\ac{DL}, \ac{ML}, and \ac{AI} have revolutionized \ac{CI} segmentation, enabling precise identification and analysis of cochlear structures in various imaging modalities. Li et al. \cite{li2023application} applied a UNETR model to automatically segment cochlear structures in temporal bone \ac{CT} images, enhancing surgical planning and cochlear implantation outcomes. Reda et al. \cite{reda2014automatic} developed an automatic segmentation method for intra-cochlear anatomy in post-implantation \ac{CT} scans, facilitating the customization of sound processing strategies for individual \ac{CI} recipients. Moudgalya et al. \cite{moudgalya2020deep} employed a modified V-Net \ac{CNN} to segment cochlear compartments in $\mu$\ac{CT} images, enabling precise quantification of local drug delivery for potential treatment of sensorineural hearing loss. Wang et al. \cite{wang2020metal} focused on metal artifact reduction and intra-cochlear anatomy segmentation in \ac{CT} images using a multi-resolution multi-task deep network, benefiting \ac{CI} recipients. Heutink et al. \cite{heutink2020multi} developed a \ac{DL} framework for the automatic segmentation and analysis of cochlear structures in ultra-high-resolution \ac{CT} images, providing accurate measurements for personalized implant planning in cochlear surgery. Zhang et al. \cite{zhang2019two} utilized a 3D U-Net \ac{DL} method to achieve accurate segmentation of intra-cochlear anatomy in head \ac{CT} images, facilitating optimal programming of \acp{CI} and improving hearing outcomes. These studies highlight the significant impact of \ac{DL}, \ac{ML}, and \ac{AI} in advancing \ac{CI} segmentation, ultimately leading to improved patient care and treatment outcomes. 
{Recently, Zhu et al. \cite{zhu2024uadsn} proposes an uncertainty-aware dual-stream network, called UADSN, for facial nerve segmentation in \ac{CT} scans for cochlear implantation surgery. UADSN combines 2D and 3D segmentation streams and uses consistency loss to improve accuracy in uncertain regions. The network achieves superior performance compared to other methods on a facial nerve dataset, with an emphasis on topology preservation.} 

%{Recently, Zhang et al.  \cite{zhang2024monocular} proposes a method for monocular microscope to \ac{CT} registration using pose estimation of the incus for augmented reality \ac{CI} surgery. The proposed solution achieves direct 2D-to-3D registration without external tracking equipment. Results show an average rotation error of less than 25 degrees and a translation error of less than 2 mm, 3 mm, and 0.55\% for the x, y, and z axes, respectively.}

\subsection{Thresholding}

\ac{DL}, \ac{ML}, and \ac{AI} have been instrumental in the field of \acp{CI}, particularly in thresholding applications. Kuczapski et al. \cite{kuczapski2015assistive} developed a software tool that utilizes \ac{AI} to estimate and monitor the \ac{EST} levels in \ac{CI} recipients. By leveraging patient data, audiograms, and fitting settings, this tool aids in the fitting process and predicts changes in hearing levels, enhancing personalized care. Botros et al. \cite{botros2007autonrt} introduced AutoNRT, an automated system that combines \ac{ML} and pattern recognition to measure \ac{ECAP} thresholds with the Nucleus Freedom \ac{CI}. This objective fitting system streamlines clinical procedures and ensures precise and efficient threshold measurements. Furthermore, Schuerch et al. \cite{schuerch2023objective} utilized a \ac{DL}-based algorithm to objectively evaluate and analyze \ac{ECochG} signals. This algorithm enables the assessment of \ac{ECochG} measurement repeatability, comparison with audiometric thresholds, and identification of signal patterns and tonotopic behavior in \ac{CI} recipients. Through the integration of \ac{DL}, machine \ac{ML}, and \ac{AI}, these studies have significantly advanced thresholding techniques in \ac{CI} applications, leading to improved fitting accuracy, streamlined procedures, and objective evaluation of signal responses.

\subsection{Localization of \ac{CI} }
\ac{DL} methods have been instrumental in \ac{CI} localization applications, providing accurate and automated solutions. Chi et al. \cite{chi2019deep}
proposed a \ac{DL}-based method for precise localization of  electrode contacts in \ac{CT} images. Their approach utilized cGANs to
generate likelihood maps, which were then processed to estimate the exact location of each contact. Radutoiu et al. \cite{radutoiu2022accurate} focused
on the automatic extraction of \acp{ROI} in full head \ac{CT} scans of the inner ear. By leveraging \ac{AI}, they achieved high precision in \ac{ROI}
localization, facilitating accurate surgical planning for  insertion. Noble et al. \cite{noble2015automatic} and Zhao et al. \cite{zhao2014automatic,zhao2019automatic}, developed AI-based
systems to automatically identify and position electrode arrays in \ac{CT} images. These technologies enable large-scale analyses of the
relationship between electrode placement and hearing outcomes, leading to potential advancements in implant design and surgical
techniques. Heutink et al. \cite{heutink2020multi} employed \ac{DL} for the automatic segmentation and localization of the cochlea in ultra-high-resolution
\ac{CT} images. This approach allows for precise measurements that can be used in personalized  planning, reducing the risk of
intra-cochlear trauma and optimizing surgical outcomes. These studies showcase the significant contributions of \ac{DL} and \ac{AI} in 
localization applications, enabling accurate and efficient identification, positioning, and analysis of electrode arrays and facilitating
improved surgical planning and outcomes. {Burkart et al. \cite{burkart2024analysis} investigates the influence of sound source position and electrode placement on the stimulation patterns of \ac{CI} under noise conditions. The study utilizes a measurement setup to simulate realistic listening scenarios. The results reveal that the effectiveness of \ac{CI} noise reduction systems is influenced by these factors, and artificial intelligence fitting algorithms should be considered to optimize \ac{CI} performance.
}

\subsection{Other}

\ac{DL} techniques have been employed in various \ac{CI} applications, showcasing their potential to enhance hearing outcomes and improve device performance. Bermejo et al. \cite{bermejo2013probabilistic} introduced a decision support system using a novel probabilistic graphical model to optimize \ac{CI} parameters based on audiological tests and the current device status, aiming to optimize the user's hearing ability. Castaneda et al. \cite{castaneda2007objective} focused on the use of \ac{BSS} and independent component analysis (ICA) to identify \acp{AEP} and isolate artifacts in children with \acp{CI}, enabling improved assessment of auditory function. Incerti et al. \cite{incerti2019effect} investigated the impact of varying cross-over frequency settings for \ac{EAS} on binaural speech perception, localization, and functional performance in adults with \acp{CI} and residual hearing, providing valuable insights for personalized device programming. Katthi et al. \cite{katthi2020deep} developed a \ac{DL} framework based on \ac{CCA} to decode the auditory brain, establishing a strong correlation between audio input and brain activity measured through \ac{EEG} recordings. This research has implications for decoding human auditory attention and improving \acp{CI} by leveraging the power of \ac{DL}. 

\section{Open issues and future directions}
\label{sec5}

While significant strides have been achieved in integrating \ac{AI} into  \acp{CI}, numerous research lacunae persist, offering avenues for further advancements in the field. Here are several potential realms warranting exploration in future studies:\\

\noindent\textbf{Real-time signal processing and 
 personalized design: } Investigating real-time adaptive signal processing methods employing \ac{AI} algorithms has the potential to enhance sound processing for \ac{CI} recipients, yielding enhanced speech intelligibility outcomes. Enhancements in adaptability to dynamic acoustic environments and real-time optimization of stimulation parameters have the capacity to substantially enhance \ac{CI} performance. The authors have observed a gap in the study and implementation of \ac{AI} models tailored for \ac{CI} on real-time platforms like field programmable gate arrays (FPGAs). Further research in this burgeoning area holds promise for adapting a variety of existing AI models to enhance real-time capabilities.
 
 \vspace{0.2cm}
 Besides, tailoring \acp{CI} to meet the unique needs of individual users poses a significant challenge. Investigating \ac{AI}-driven methodologies leveraging personal data to personalize device configurations based on factors like physiological, auditory, and neural feedback during mobility can enhance both individual outcomes and overall satisfaction.
 
\vspace{0.2cm}
\noindent\textbf{Predicting the long-term effects:} Gaining insight into the enduring \acp{CI} is vital for enhancing patient selection, counseling, and device advancement. Utilizing \ac{AI} methods to shift through extensive datasets can pinpoint the predictive elements influencing sustained success. These factors may encompass pre-implantation attributes, surgical approaches, and auditory rehabilitation. Constructing predictive models using AI algorithms can furnish valuable perspectives on long-term consequences, thereby informing clinical judgments.
\vspace{0.2cm}

\noindent\textbf{Incorporating multiple sensory and modalities:} \acp{CI} traditionally prioritize the reinstatement of auditory experiences. Yet, enriching the perception and comprehension of sound can be achieved by integrating additional sensory dimensions like vision and touch, resulting in a multi-modal approach. Investigating \ac{AI}-driven techniques that amalgamate inputs from various senses to enhance speech recognition, spatial sound perception, and overall auditory understanding presents a promising direction for future exploration. Besides, \acp{CI} in both ears, when paired with \ac{AI} algorithms, can enhance speech comprehension. By analyzing sound patterns from both implants, AI adjusts settings to optimize signal processing, improving overall accuracy and clarity of speech perception for users with bilateral implants, and enhancing their auditory experience and communication abilities, as investigated in \cite{gajecki2024fused}. However, extensive research possibilities are required to tailor solutions with \ac{CI} hardware capabilities, by taking into account computation cost, and \ac{AI} model complexity.
\vspace{0.2cm}

\noindent\textbf{Empowering AI-based CI using DTL:} \Ac{DTL} is a highly efficient \ac{DL} technique enabling the transfer of knowledge from pre-trained models, trained on millions of speech corpora and/or images, to train smaller models with limited data availability \cite{himeur2023video,kheddar2023deepIDS}. This approach offers significant advantages in producing lightweight \ac{AI} models suitable for devices with limited computational resources, such as \acp{CI}. Only a limited number of studies have explored the impact of \ac{DTL} on \ac{CI}, as demonstrated in \cite{li2021improved}, which has received relatively little attention from researchers. We anticipate further exploration of this promising technique, particularly through the utilization of various \ac{DTL} sub-techniques, such as domain adaptation, transductive methods like cross-lingual transfer, cross-corpus transfer, zero-shot learning, fine-tuning, among others \cite{KheddarASR2023}. 

\vspace{0.2cm}
\noindent\textbf{Ensuring data privacy through FL:}
\Ac{FL} facilitates collaborative model training across decentralized devices by aggregating local updates rather than centralizing data. This preserves user privacy and enhances model performance, particularly beneficial in healthcare applications \cite{himeur2023federated}. Gathering comprehensive datasets is challenging due to rare anomaly cases and privacy concerns. \ac{FL} addresses this by training models on distributed, encrypted data from multiple sources, ensuring privacy while maintaining efficacy. Researchers have yet to fully explore FL-based model building for \ac{CI}, neglecting the potential to construct efficient \ac{AI} models capable of accommodating diverse classes. Further investigation into this promising technique is warranted, with potential for significant advancements in model robustness and versatility. Moreover, this approach could lead to the development of a pretrained model utilizing \ac{FL}, which could be seamlessly integrated with \ac{DTL}.

\vspace{0.2cm}

\noindent\textbf{Transformers-based CI techniques: } AI researchers have adopted the CNN-LSTM model that excels in capturing spatial and temporal features, enhancing performance in sequential data tasks \cite{djeffal2023noise,gueriani2024enhancing,kheddar2024transformers}. However, Transformer-based  ASR techniques, such as \ac{CTC}, \ac{BERT}, and others, proved in the literature that have the potential to greatly enhance the functioning of \ac{ASR} \cite{djeffalTr2024,kheddar2024automatic,habchi2024machine}. By leveraging the self-attention mechanism, transformers can improve speech intelligibility by effectively suppressing background noise and modeling long-range dependencies. They can also aid in acoustic scene analysis, separating and prioritizing important auditory information in complex environments. Transformers can build language models that enhance \ac{ASR} systems, improving speech comprehension for users. Additionally, transformers enable personalized sound processing by adapting stimulation patterns and processing parameters based on user-specific preferences. They facilitate multi-modal integration, combining audio and visual inputs to enhance speech recognition and sound localization. Furthermore, transformers support long-term learning and adaptation, continually optimizing \ac{CI} performance over time. These advancements offer promising prospects for improving auditory experiences and overall quality of life for \ac{CI} users.

\vspace{0.2cm}
\noindent\textbf{Exploring chat-bots-based CI capabilities: }
Chat-bot techniques offer several opportunities to enhance the functioning of \acp{CI}. They can provide real-time support, troubleshooting, and personalized rehabilitation programs for users, empowering them to address common issues and improve their auditory skills. Chat-bots enable remote monitoring, allowing users to share data and receive adjustments to their device settings without in-person appointments. They also offer emotional and psychological support, fostering a sense of community and well-being. Chat-bots contribute to data collection for research and development, aiding in the improvement of \ac{CI} technology and rehabilitation protocols. Additionally, chat-bots employ\ac{ML} to continuously learn from user interactions, improving their responses and understanding over time. These techniques have the potential to enhance the overall user experience, outcomes, and accessibility of \ac{CI} services. {For example, in \cite{aliyeva2024enhancing}, the effectiveness of ChatGPT-4 in providing postoperative care information to \ac{CI} patients was evaluated. Five common questions were posed to ChatGPT-4, and its responses were analyzed for accuracy, response time, clarity, and relevance. The results showed that ChatGPT-4 provided accurate and timely responses, making it a reliable supplementary resource for patients in need of information.}

\section{Conclusion}
\label{sec6}

This review has provided a comprehensive overview of the advancements in \ac{AI} algorithms for \acp{CI} applications and their impact on \ac{ASR} and speech enhancement. The integration of \ac{AI} methods has brought cutting-edge strategies to address the limitations and challenges faced by traditional signal processing techniques in the context of \acp{CI}. Moreover, the application of \ac{AI} in \ac{CI} has led to the emergence of new datasets and evaluation metrics, offering alternative methods for validating proposed schemes without the need for human surgical intervention and traditional tests. The review highlighted the role of \ac{ASR} in optimizing speech perception and understanding for \ac{CI} users, contributing to the improvement of their quality of life. \ac{ASR} not only enhances basic speech recognition but also aids in the recognition of environmental sounds, enabling a more immersive auditory experience. 
Furthermore, \ac{ASR} finds applications in authentication systems, event recognition, source separation, and speaker recognition, extending its reach beyond communication.
Various \ac{AI} algorithms, belong to \ac{ML} and  \ac{DL}, have been explored in the context of \acp{CI}, demonstrating promising results in speech synthesis and noise reduction. These algorithms have shown the potential to overcome challenges associated with multiple sources of speech, environmental noise, and other complex scenarios. The review has summarized and commented on the best results obtained, providing valuable insights into the capabilities of \ac{AI} algorithms in this biomedical field.
Moving forward, the review suggests future directions to bridge existing research gaps in the domain of \ac{AI} algorithms for \acp{CI}. It emphasizes the need for high-quality data inputs, algorithm transparency, and collaboration between researchers, clinicians, and industry experts. Addressing these aspects will facilitate the development of more accurate and efficient \ac{AI} algorithms for \ac{CI}, ultimately benefiting individuals with hearing impairments. The integration of advanced \ac{AI} algorithms has the potential to revolutionize the field of \acp{CI}, providing individuals with hearing impairments to better communicate and engage with the world around them. Continued research and development in this area hold great promise for the future of \ac{CI} technology.

%\section*{Acknowledgement}
%The second author acknowledges that the study was partially funded by the Algerian Ministry of Higher Education and Scientific Research (Grant No. PRFU--A25N01UN260120230001).

\section*{Data availability}
Data will be made available on request.

\section*{Conflict of Interest}
The authors declare no conflicts of interest.

\balance
\bibliographystyle{IEEEtran}
\bibliography{references.bib}
\begin{IEEEbiography}[{\includegraphics[width=1in,height=1.25in,clip,keepaspectratio]{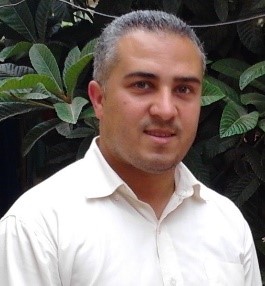}}]{Billel Essaid} 
is an accomplished Algerian researcher and lecturer specializing in telecommunications and biomedical engineering. He currently holds the position of Assistant Professor in the Department of Electrical Engineering at the University of Médéa in Algeria. With a strong educational background and a passion for scientific exploration, he has dedicated his career to advancing knowledge in his fields of expertise. He received a bachelor's degree in 1999 and an engineering degree in Electronics in 2004, specializing in electronic instrumentation from Médéa University in Algeria. He then earned a Magister degree in Electronic Systems with a focus on telecommunications from the Military Polytechnic School in Algeria in 2007. His academic achievements have paved the way for a successful career in academia, where he actively contributes to the growth and development of the next generation of engineers. Throughout his career, he has supervised numerous Master's projects, covering areas such as cochlear implant stimulation, noise reduction, and speech signal processing.
\end{IEEEbiography}

\begin{IEEEbiography}[{\includegraphics[width=1.in,height=1.05in,clip,keepaspectratio]{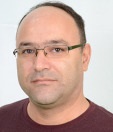}}]{HAMZA KHEDDAR} 
senior IEEE member, is currently an Associate Professor at the University of Medea and Researcher at LSEA Lab, Medea (Algeria), holds a Ph.D. in Telecommunication from USTHB University. With expertise in speech steganography, digital watermarking, Telecommunication, Intrusion detection, and more, he obtained the Habilitation to Direct Research in June 2021. Serving as a reviewer for esteemed journals like \textit{Computers \& Security}, \textit{IEEE Access}, \textit{IEEE Transactions forensics information security}, \textit{ IEEE communications letters},  Computer Speech \& Language, Applied intelligence, speech communication and others. He contributes significantly to diverse research areas, with more than 30 paper, such as image classification, intrusion detection, artificial intelligence, 5G/6G slicing and security. He is also a telecommunication track chair in many conferences, such as IC2M 2023, and ICTSS 2024. 
\end{IEEEbiography}

\begin{IEEEbiography}[{\includegraphics[width=1.in,height=1.25in,clip,keepaspectratio]{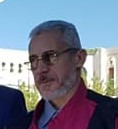}}]{Noureddine Batel} 
He is the head of the Microelectronics and Real-Time Information Processing group at the LSEA Laboratory, and a distinguished researcher in the field of Electronics. He obtained his Doctorate in Electrical Engineering from the National Polytechnic School in Algeria in 2007 and served as the former dean of the Faculty of Technology at the University of Médéa. His research interests include signal and speech processing, electronic circuits, and telecommunications. Currently, he holds the position of Professor in the Department of Electrical Engineering at the University of Médéa in Algeria.
\end{IEEEbiography}

\begin{IEEEbiography}[{\includegraphics[width=1in,height=1.25in,clip,keepaspectratio]{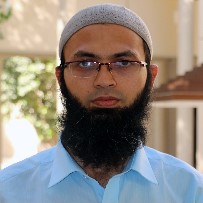}}]{MUHAMMAD E. H. CHOWDHURY} 
received his Ph.D. degree from the University of Nottingham, U.K., in 2014. He worked as a Postdoctoral Research Fellow at the Sir Peter Mansfield Imaging Centre, University of Nottingham. He is currently working as an Assistant Professor and program coordinator of the Department of Electrical Engineering, Qatar University. He has filed several patents and published more than 200 peer-reviewed journal articles, 30+ conference papers, and several book chapters. His current research interests include biomedical instrumentation, signal processing, wearable sensors, medical image analysis, machine learning and computer vision, embedded system design, and simultaneous EEG/fMRI. He is currently running NPRP, UREP, and HSREP grants from Qatar National Research Fund (QNRF) and internal grants (IRCC and HIG) from Qatar University along with academic projects from HBKU and HMC. He is a Senior Member of IEEE, and a member of British Radiology, ISMRM, and HBM. He is serving as Guest Editor for Polymers, an Associate Editor for IEEE Access and a Topic Editor and Review Editor for Frontiers in Neuroscience. He has recently won the COVID-19 Dataset Award, AHS Award from HMC and National AI Competition awards for his contribution to the fight against COVID-19. His team is the gold-medalist in the 13th International Invention Fair in the Middle East (IIFME). He has been listed among the Top 2\% of scientists in the World List, published by Stanford University. 
\end{IEEEbiography}

\begin{IEEEbiography}[{\includegraphics[width=1in,height=1.25in,clip,keepaspectratio]{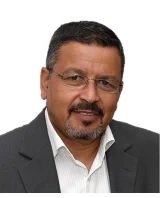}}]{Abderrahmane Lakas}  Abderrahmane Lakas (SMIEEE) is a Professor of Computer and Network Engineering at the College of
Information Technology (CIT), UAE University (UAE). He holds a PhD and MSc in Computer Systems from
the Pierre and Marie Curie University, France, and a BSc in Computer Systems from the National
Institute of Informatics (Currently ESI, Algeria). He has several years of both academic and industrial
experience. Before joining UAE University, Prof. Lakas worked in the telecommunications industry in
various technical and managerial positions. His current research interests cover a wide range of areas
including vehicular ad-hoc networks (VANET), unmanned aerial vehicles (UAV), the Internet of Things
(IoT), Internet of vehicles (IoV), edge computing, wireless communication (3GPP, 5G, 6G), machine
learning, deep reinforcement learning, game theory, brain-computer interfaces (BCI), Quality of Service
(QoS), Peer-to-Peer (P2P) networks, and cybersecurity. He is currently the head of CAST (Connected
Autonomous Intelligent Systems) Lab. He is the author or co-author of more than a hundred refereed
journal and conference papers. His research has been widely published in reputable journals such as
Vehicular Communication Journal, IEEE Transactions on Vehicular Technology, IEEE Internet of Things
Journal, IEEE Transactions on Intelligent Transportation Systems, ACM Computing Surveys, etc. He
actively participates in conferences such as IEEE VTC, IEEE CC, IEEE Globcom, IEEE WCNC. He is a senior
member of IEEE and serves as an Associate Editor for IEEE Access, IET Networks, and has held past
editorial roles in the Journal of Computer Systems, Networks, and Communications and the
International Journal of Communications. Notably, his research has been recognized with several awards
such as best paper awards in conferences and journals. Prof. Lakas appears in the prestigious
Stanford/Elsevier's list of the top 2\% of world scientists.
\end{IEEEbiography}

\EOD

\end{document}